\documentclass[11pt]{article}
\usepackage{amsmath}
\usepackage{amsfonts}
\usepackage{hyperref}
\usepackage{epsfig}
\newcommand{\ket}[1]{|{#1}\rangle}

\begin{document}

\author{John Preskill}

\title{Quantum computing 40 years later}
\author{John Preskill\\ \\ \sl{Institute for Quantum Information and Matter} \\ \sl{California Institute of Technology, Pasadena CA 91125, USA}\\ \sl{AWS Center for Quantum Computing, Pasadena CA 91125, USA}}
\date{6 June 2021}
\maketitle

\begin{abstract}
Forty years ago, Richard Feynman proposed harnessing quantum physics to build a more powerful kind of computer. Realizing Feynman's vision is one of the grand challenges facing 21st century science and technology. In this article, we'll recall Feynman's contribution that launched the quest for a quantum computer, and assess where the field stands 40 years later. To appear in \emph{Feynman Lectures on Computation, 2nd edition}, published by Taylor \& Francis Group, edited by Anthony J.~G.~Hey.
\end{abstract}

\maketitle
\tableofcontents
\vspace{1cm}

Forty years ago, Richard Feynman proposed harnessing quantum physics to build a more powerful kind of computer. Realizing Feynman's vision is one of the grand challenges facing 21st century science and technology. In this article, we'll recall Feynman's contribution that launched the quest for a quantum computer, and assess where the field stands 40 years later. 

After a brisk account in Sec.~\ref{sec:feynman-qc} and \ref{sec:where} of how quantum computing has developed over the past few decades, I sketch the foundations of the subject, discussing salient features of quantum information in Sec.~\ref{sec:information}, then formulating a mathematical model of quantum computation in Sec.~\ref{sec:computer} and highlighting some implications of the model. In Sec.~\ref{sec:time-evolution} and Sec.~\ref{sec:energy}, I review two particularly promising applications of quantum computing foreseen by Feynman, simulating the dynamics of complex quantum systems, and computing their static properties. In Sec.~\ref{sec:qec}, I explain the concept of quantum error correction, the basis of our belief that quantum computers can be scaled up to large systems that solve very hard problems. I offer some concluding thoughts in Sec.~\ref{sec:outlook}, and reminisce about some of my interactions with Feynman in Sec.~\ref{sec:memories}.

\section{Feynman and quantum computation}
\label{sec:feynman-qc}

\subsection{Feynman's 1981 talk}
Richard Feynman was renowned for his wide ranging intellect, and computation was one of the many topics that fascinated him. He was interested in the scientific applications of computing, but also deeply curious about how hardware and software really work, and about fundamental issues in the theory of computation. These interests are amply reflected in \textit{The Feynman Lectures on Computation}.

In May 1981, Feynman spoke at a conference on the topic ``Simulating physics with computers.'' There he proposed the idea of using quantum computers to simulate quantum systems that are too hard to simulate using conventional classical digital computers. Feynman's talk, later published as a lightly edited transcript \cite{feynman21simulating}, is justly remembered for its role in launching quantum computing as a field of study.

In the talk, Feynman clearly states his goal, which is to simulate quantum systems using resources that scale well with the size of the system:

\begin{quote}
The rule of simulation that I would like to have is that the number of computer elements required to simulate a large physical system is only to be proportional to the space-time volume of the physical system.
\end{quote}

\noindent He points out why digital computers are not adequate for the task, because there is no succinct way to describe classically a quantum state of many particles:

\begin{quote}
Now I explicitly go to the question of how we can simulate with a computer \dots the quantum mechanical effects \dots But the full description of quantum mechanics for a large system with $R$ particles is given by a function which we call the amplitude to find the particles at $x_1, x_2, \dots x_R$, and therefore because it has too many variables, \emph{it cannot be simulated with a normal computer} (italics mine).
\end{quote}
\noindent In a crucial passage, he speculates that a different kind of computer might be up to the job:

\begin{quote}
Can you do it with a new kind of computer --- a quantum computer? Now it turns out, as far as I can tell, that you can simulate this with a quantum system, with quantum computer elements. \emph{It’s not a Turing machine, but a machine of a different kind} (italics mine).
\end{quote} 
\noindent And Feynman challenges the computer scientists to study the power of this new model of computation:

\begin{quote}
I present that as another interesting problem: To work out the classes of different kinds of quantum mechanical systems which are really intersimulatable --- which are equivalent --- as has been done in the case of classical computers.
\end{quote}

About half of the talk is devoted to elaborating the argument that digital computers will be inadequate for efficiently simulating quantum systems. He emphasizes that quantum theory will not admit a local hidden variable description, and there follows a lucid discussion of Bell inequalities and the experimental evidence that these are violated (without any references and without ever mentioning Bell). 
\begin{quote}
If you take the computer to be the classical kind I’ve described so far (not the quantum kind described in the last section) and there’re no changes in any laws, and there’s no hocus-pocus, the answer is certainly, “No!” This is called the hidden variable problem: It is impossible to represent the results of quantum mechanics with a classical universal device.
\end{quote} 

\noindent Perhaps most famously, Feynman brought the talk to a stirring conclusion:
\begin{quote}
Nature isn’t classical, dammit, and if you want to make a simulation of Nature, you’d better make it quantum mechanical, and by golly it’s a wonderful problem because it doesn’t look so easy.
\end{quote}

\noindent Feynman, then nearly 63, was calling for a fundamentally new type of computing machine, and he foresaw its natural application: making ``a simulation of Nature.'' Bringing that vision to fruition is indeed ``a wonderful problem'' and 40 years later it still ``doesn't look so easy.''

\subsection{Manin and Benioff}

Around the same time, a few others were considering quantum models of computation, including the distinguished mathematician Yuri Manin. 
In his 1980 book \emph{Computable and Uncomputable} \cite{manin1980computable}, Manin, like Feynman, cogently emphasized the exponential cost of simulating a many-particle particle system with a classical computer. Manin wrote [translated from the Russian by Victor Albert]:

\begin{quote}
These objects [quantum automata] may show us mathematical models of deterministic processes with highly unusual features. One of the reasons for this is because \emph{the quantum phase space is much bigger than classical}: where classical space has $N$ discrete levels, a quantum system allowing their superposition will have $c^N$ Planck cells. In a union of two classical systems, their sizes $N_1$ and $N_2$ multiply, but in the quantum case we have $c^{N_1+N_2}$.
These heuristic calculations point to a much larger potential complexity of the behavior of a quantum system when compared to its classical imitator 
(italics mine).
\end{quote}

\noindent Also in 1980, Paul Benioff explained how to describe computation from a fundamentally quantum point of view \cite{benioff1980computer}. He wrote:

\begin{quote}
These considerations suggest that it may be impossible even in principle to construct a quantum mechanical Hamiltonian model of the computation process. The reason is that any such model evolves as an isolated system with a constant total energy. The point of this paper is to suggest, by construction of such models, that this may not be the case.
\end{quote}

\noindent Unlike Manin and Feynman, Benioff was not concerned with quantum \emph{complexity}. Rather, he mainly focused on the question whether a quantum computer can operate without dissipation.  Feynman was also deeply interested in this issue, and addressed it in detail in his talk at the 1984 CLEO/IQEC conference (``Quantum Mechanical Computers''), which is reprinted in this volume.

Strangely, though Feynman broached the topic of quantum computing in the lecture course that is captured by this book, he never mentioned in his class lectures the deep idea, so clearly articulated in his 1981 talk, that quantum computers can solve certain problems far more efficiently than classical computers. Yet what could be a better fit to a course on ``The Potentialities and Limitations of Computing Machines''? I find this omission baffling. 

\subsection{From Feynman to Shor and beyond}

It took a while, but gradually the influence of Feynman's ideas grew. In 1985, David Deutsch formalized the notion of a quantum computer \cite{deutsch1985quantum}, an important advance, and raised the question whether quantum computers might have an advantage over classical computer at solving problems that have nothing to do with quantum physics. In 1993, Umesh Vazirani and his student Ethan Bernstein formulated a contrived problem that a quantum computer could solve with a superpolynomial speedup over a classical computer \cite{bernstein1997quantum}. Soon after, Daniel Simon showed that a quantum computer could achieve an exponential speedup in solving an idealized version of the problem of finding the period of function \cite{simon1997power}. Though Simon's problem had no obvious applications, it inspired Peter Shor \cite{shor1999polynomial}, who worked out a very efficient way of performing a Fourier transform using a quantum computer, and applied it to formulate an efficient quantum algorithm for computing discrete logarithms. Just days later, Shor used similar ideas to find an efficient quantum algorithm for factoring large numbers.

Shor's discovery, and its obvious implications for cryptanalysis, caused interest in quantum computing to skyrocket. But very good physicists like Rolf Landauer \cite{landauer1995quantum}, Bill Unruh \cite{unruh1995maintaining}, and Serge Haroche \cite {haroche1996quantum} voiced strong skepticism about whether quantum computers could ever work effectively. Those physicists were deeply familiar with the debilitating effects of \emph{decoherence}, which under typical conditions prevent complex many-particle quantum systems from exhibiting quantum behavior, and viewed quantum computing (in the derisive words of Haroche and Raimond) as ``the computer scientist's dream [but] the experimenter's nightmare.'' Remarkably, it was again Shor who led the next pivotal advances --- the discovery of quantum error-correcting codes \cite{shor1995scheme,steane1996error} and of fault-tolerant methods for executing a quantum computation reliably using noisy hardware \cite{shor1996fault}. By the end of 1996, it was understood at least in principle that quantum computing could be scaled up to large devices that solve very hard problems, assuming that errors afflicting the hardware are not too common or too strongly correlated \cite{aharonov1997fault,knill1998resilient,kitaev1997quantum,preskill1998reliable,preskill1998fault}. This ``accuracy threshold theorem'' for quantum computing was already in place 2$\frac{1}{2}$ years after the discovery of Shor's algorithm. Meanwhile, Cirac and Zoller had proposed that tools in atomic physics and quantum optics can be exploited to perform quantum logical operations \cite{cirac1995quantum}, amplifying interest in the experimental physics community \cite{monroe1995demonstration}. Those were exciting times \cite{preskill1998quantum}. 

\subsection{Imagining the future}

We don't know exactly how Feynman arrived at the idea of a quantum computer, but we do know that by 1981 he had been thinking about the future of computing for decades. In his remarkable 1959 talk called ``There’s Plenty of Room at the Bottom'' \cite{feynman1960there}, which foresaw the field of nanotechnology, he mused about the miniaturization of computing circuitry:

\begin{quote}
If they had millions of times as many elements, they could make judgements \dots In many ways, they would have new qualitative features \dots  There is nothing that I can see in the physical laws that says computer elements cannot be made enormously smaller than they are now. In fact, there may be certain advantages. 
\end{quote}

\noindent And he imagined devices assembled by precise placement of single atoms:

\begin{quote}
All our devices can be mass produced so that they are absolutely perfect copies of one another \dots The principles of physics, as far as I can see, do not speak against the possibility of maneuvering things atom by atom.
\end{quote}

\noindent That willingness to look beyond the limitations of current technology and envision what might be possible in the future is equally apparent in his speculations about ``simulating physics with computers.'' To computer scientists of the early 1980s, quantum physics was viewed as an obstacle that would ultimately block further miniaturization of electrical circuitry. But to Feynman, quantum physics offered a dazzling opportunity. As his lectures on computing attest, Feynman knew enough about the theory of computation to understand and appreciate the ``extended Church-Turing thesis,'' which asserts that any physically realizable universal computer can efficiently simulate any other (under a loose notion of ``efficiently''). The great insight trumpeted in his 1981 talk is that this story needs revision because ``Nature isn't classical, dammit.'' 
That idea could change the world.

\section{Where we're going and where we are}
\label{sec:where}

\subsection{How will quantum computers be used?}

Because of Shor's algorithm, the public key cryptographic protocols we use to protect our privacy when we communicate over the Internet will become vulnerable to attacks by quantum computers in the future. To protect against that threat, ``quantum-resistant'' protocols are now being developed, based on computational problems that we think are too hard to solve even with a quantum computer \cite{bernstein2017post}. An alternative path is quantum cryptography, where quantum states are exchanged over a quantum communication network, and security rests on the principle that eavesdropping on quantum communication produces an unavoidable and detectable disturbance. (That is another fascinating story of quantum technology \cite{wiesner1983conjugate,bennett2020quantum}, which unfortunately is beyond the scope of this article.) Perhaps both of these approaches will be deployed, depending on the needs of the user \cite{mosca2018cybersecurity}. 

Quantum computing is such a big departure from previous methods for processing information that no one can be expected to foresee its long-term ramifications. But based on what we currently understand about the power of quantum computing, Feynman's proposal to use quantum computers to simulate quantum systems is still the application that seems most likely to have a broad impact on the world. More powerful methods in computational quantum chemistry, for example, may eventually yield significant improvements in human health (pharmaceuticals), agriculture (nitrogen fixation), and the sustainability of our planet (energy storage and production, carbon sequestration) \cite{mcardle2020quantum}. In contrast, while Shor's algorithm for factoring will have a disruptive effect on electronic commerce in the relatively near term, its long-term implications are not likely to be comparably profound. 

It is important to emphasize that quantum computers have limitations. We don't expect, in particular, that quantum computers can find exact solutions to NP-hard optimization problems efficiently \cite{bennett1997strengths}. There is a general scheme for speeding up exhaustive search for a solution using quantum computers (Grover's algorithm), but the speed-up is quadratic in that case \cite{grover1997quantum}; that is, the quantum computer finds the solution in a time that scales as the square root of the time needed by the classical computer. Under the highly idealized assumption that the classical and quantum computers have the same clock speed (can perform the same number of elementary operations per second), that means that the quantum computer can find a solution that is $2n$ bits long in the time it would take the classical computer to find a solution that is $n$ bits long, for asymptotically large $n$. That may be important someday. But for factoring large numbers or simulating quantum systems the speed-up is far more spectacular. The runtime for simulating an $n$-qubit quantum system using a classical computer, in the hardest instances, rises exponentially with $n$, while the runtime for simulating the system on a quantum computer scales like a power of $n$. That is a game-changing difference, as Feynman appreciated.

\subsection{The NISQ era unfolds}

It is also important to emphasize that quantum computers are not very useful yet. In the 40 years since Feynman's talk, a variety of approaches to building quantum hardware have emerged and progressed, but both the number of qubits and the accuracy of our quantum processors are still quite modest. An indicator of the current status is a milestone declaimed in 2019 by the Google AI Quantum group \cite{arute2019quantum}, known as ``quantum computational supremacy'' \cite{preskill2012quantum,harrow2017quantum}.

As Feynman emphasized, a remarkable claim about quantum physics is that classical systems cannot in general simulate quantum systems efficiently. Arguably that is one of the deepest statements known about the difference between quantum and classical, and we have a strong incentive to validate it in the laboratory to the extent that we can. Can we exhibit some task performed by a quantum computer that would require a much longer runtime on any existing classical computer? 

Using superconducting quantum technology, the Google group constructed a programmable quantum computer called Sycamore with 53 working qubits arranged in a two-dimensional array, such that entangling two-qubit quantum gates can be performed on neighboring qubits in the array. (We will get to an explanation of what ``entangling'' means in Sec.~\ref{sec:information}.) They executed up to 20 layers of two-qubit gates, and then measured all the qubits at the end. Because the hardware sometimes makes errors, the final measurement yields the correct output only once in 500 runs of the computation. But by repeating the same computation millions of times in just a few minutes, they extracted a statistically useful signal. 

Using the best currently known methods, simulating what Sycamore does in a few minutes would take at least a few days for the most powerful existing classical supercomputer \cite{huang2020classical}. Furthermore, the cost of the classical simulation rises exponentially with the number of qubits, and so would be very far beyond the classical computer's reach if only a few more qubits could be added. While the classical computer occupies the equivalent of two tennis courts and consumes megawatts of power, Sycamore is just a single chip nested inside a dilution refrigerator. Quantum David overpowers Classical Goliath. 

Admittedly, this task performed by Sycamore is of no particular interest for any purpose besides demonstrating quantum computational supremacy. But it signals that quantum hardware is now working well enough to produce meaningful results in a regime where classical simulation is very difficult, and motivates us to seek more meaningful applications. 

It is convenient to have a word for the new quantum era that is now opening, and the word NISQ has caught on \cite{preskill2018quantum}. It stands for \emph{Noisy Intermediate-Scale Quantum}. ``Intermediate scale'' conveys that today's quantum devices with more than 50 well controlled qubits cannot be simulated by brute force using the most powerful currently existing classical supercomputers; ``noisy'' reminds us that these devices are not error corrected, and that the noise limits their computational power. For physicists, NISQ technology is exciting --- it provides us with new tools for exploring the properties of highly complex many-particle quantum systems in a regime which has never been experimentally accessible before. It might also have other applications of interest to a broader community of users, but we're not yet sure about that. NISQ will not change the world by itself, at least not right away; instead we should regard it as a step toward more powerful quantum technologies we hope to develop in the future. 

In the most advanced multi-qubit quantum processors that are currently available, the probability that a two-qubit quantum gate makes a sizable error is slightly less than 1\%. That was why the 53-qubit Sycamore device was unable to execute circuits with more than 20 time steps. We have no convincing argument that a quantum computation with (say) of order 100 qubits and fewer than 100 time steps can solve practical problems. 

One heuristic proposal is to search for approximate solutions to optimization problems using a hybrid quantum/classical approach \cite{farhi2014quantum,peruzzo2014variational}. It makes sense to rely heavily on our powerful classical processors, and then attempt to boost that power with a NISQ co-processor. 
But we just don't know yet whether this hybrid method can outperform the best purely classical hardware running the best classical algorithms for solving the same problems. Frankly it's a lot to ask, considering that the classical methods are well honed after decades of development, and the NISQ processors are becoming available for the first time now. But we'll have to try it and see how well it works. Vibrant discussions are already underway among potential application users, hardware providers, and quantum algorithm experts. As we experiment with NISQ technology, we'll learn more about how it performs and perhaps that will point us toward promising  ideas for applications.

\subsection{Quantum simulation in the NISQ era}

Though it's not clear how we'll fare in our quest for NISQ applications to practical problems of potential commercial interest, I do feel optimistic about using quantum platforms to advance physics over the next five to ten years. Despite their notable limitations, NISQ processors should be able to prepare and study exotic quantum states that have not been accessible in the physics lab until now. 

Classical computers are especially bad at simulating \emph{quantum dynamics} --- that is, predicting how a highly-entangled quantum state will change with time. Quantum computers should have a big advantage for that task.
It is instructive to recall that the theory of classical chaos (the extreme sensitivity to initial conditions in classical dynamical systems, which accounts for our inability to predict the weather more than two weeks out) advanced rapidly in the 60s and 70s after it became possible to simulate chaotic dynamical systems using classical computers. We may anticipate that the emerging ability to simulate chaotic quantum systems (those in which entanglement spreads very rapidly) will promote advances in our understanding of quantum chaos. Valuable insights might already be gleaned using noisy devices with of order 100 qubits. 



I should comment about the distinction between analog and digital quantum simulation. When we speak of an \textit{analog quantum simulator} we mean a system with many qubits whose dynamics resembles the dynamics of a model system we are trying to study and understand. In contrast, a \textit{digital quantum simulator} is a gate-based universal quantum computer which can be used to simulate any physical system of interest when suitably programmed, and can also be used for other purposes.

Analog quantum simulation has been a very vibrant area of research for the past 20 years \cite{jaksch1998cold,greiner2002quantum}, while digital quantum simulation with general purpose circuit-based quantum computers is just now getting started. Some of the same experimental platforms, for example trapped ions and superconducting circuits, can be used for both purposes, while other systems, like trapped neutral atoms and molecules, are particularly well suited for use as analog simulators. Analog quantum simulators have been getting notably more sophisticated, and are already being employed to study quantum dynamics in regimes which may be beyond the reach of classical simulators \cite{bernien2017probing,zhang2017observation}. 
They can also be used to create highly entangled equilibrium states of quantum matter, and to study their static properties \cite{chiu2019string,mukherjee2019spectral,semeghini2021probing}.

Though analog quantum simulators are becoming increasingly programmable, they are still hampered by imperfect control --- the actual quantum system in the lab only crudely approximates the target system of interest. For that reason, analog simulators are best suited for studying features that physicists call \textit{universal}, properties which are relatively robust with respect to introducing small sources of error. A major challenge for research using analog quantum simulators is identifying accessible properties of quantum systems which are robust with respect to error, yet are also hard to simulate classically.

We can anticipate that analog quantum simulators will eventually become obsolete. Because they are hard to control, they will be surpassed someday by digital quantum simulators, which can be firmly controlled using quantum error correction. But because of the hefty overhead cost of quantum error correction, the reign of the analog quantum simulator may persist for many years. Therefore, when seeking near-term applications of quantum technology, we should not overlook the potential power of analog quantum simulators. 

In the near term, circuit-based simulations of quantum matter may be prohibitively expensive, as realistic simulations of many-particle systems will require many gates. But circuit-based methods have advantages, too, such as greater flexibility in the Hamiltonians we can study and the initial states we can prepare. Therefore, it is important to pursue both digital and analog simulation methods, keeping in mind that experience with near-term digital simulators will help to lay foundations for the more ambitious simulations we will be able to carry out in the future. The same remark applies to applications of NISQ technology more broadly.

\subsection{From NISQ to FTQC}

As I have emphasized, NISQ-era devices will not be protected by quantum error correction, and noise severely limits the scale of computations that can be executed accurately using NISQ technology. In the long run, we expect to overcome these limitations imposed by noise using quantum error correction (QEC) and fault-tolerant quantum computing (FTQC), but QEC carries a high overhead cost in the number of qubits and number of logic gates needed \cite{gottesman2010introduction,campbell2017roads}. This cost depends on which algorithms we run and on the quality of our hardware. But if the error rate per entangling two-qubit gate is, say, 0.1\% (which is better than today's hardware), we may need more than one hundred thousand physical qubits to run high-impact applications to quantum chemistry or materials science \cite{kivlichan2020improved,campbell2021early}.

It's a huge leap from where we expect to be in the next few years, with a few hundred physical qubits, to hundreds of thousands or millions of physical qubits, and that's likely to take a while. Though I'm confident that quantum computing will have a transformative impact on society eventually, that impact might still be a few decades away. No one knows for sure how long it will take to get there. Keep in mind, though, that the technology is still at an early stage, with many competing approaches, so an unanticipated breakthrough could change the outlook suddenly.

 \section{Quantum information}
\label{sec:information}
 
 Let's now probe a little deeper into what makes a quantum computer so different from an ordinary digital computer. But first, we need to understand that what a quantum processor manipulates is something different from the information processed by today's computers. 
 \subsection{Quantum vs. classical}
 What is the difference between quantum information and classical information? To a physicist, information is something we can encode and store and process in some physical system using some physical process. Since fundamentally physics is quantum mechanical,  information may be viewed as something we store  and process in a quantum state. 

For practical purposes we often get away with ignoring the nuances of quantumness. A typical macroscopic system that we might encounter in everyday life is not well isolated from its environment, and interactions with the environment continually ``measure'' the system, a phenomenon called \emph{decoherence}. A quantum system that is continually observed by its surroundings is well described by classical physics. But information carried by a quantum system (typically a microscopic one) that is sufficiently well isolated from its surroundings has intrinsic features which are not shared by classical information. Here are a few such features:

\begin{description}
\item \emph{Randomness}. Suppose a radioactive nucleus is about to emit an alpha particle. We cannot predict with certainty whether that nucleus is going to decay in the next second; we can only assign some probability, stating how likely the emission is to occur in the next second. This process is \emph{intrinsically random} in the sense that I am unable to say whether the nucleus will decay or not even if I have the most complete description of the nucleus that the laws of physics will allow. I say \emph{intrinsic} randomness to distinguish from the randomness we often encounter in everyday life, which arises from \emph{ignorance}. If I flip a coin, I know it must have come up either heads or tails, but I don't know which until I look at the coin. So, before I look, I assign probabilities to the possible outcomes, reflecting my ignorance about the true outcome. 
The intrinsic quantum randomness of alpha decay is something different. It applies even when I have the most complete possible description. 

\item \emph{Uncertainty}. When I speak of uncertainty I usually mean that the observables arising in quantum theory, the things that we can measure, don't necessarily commute. That means different possible observations can interfere with one another, not because I'm sloppy but for a fundamental reason. If two operators $A$ and $B$  don't commute, then if I measure $A$, that will unavoidably influence a measurement of $B$ that I perform afterward. In the classical world, we can in principle measure the properties of a system as accurately as we please without in any way disturbing the system. That's not true in the quantum world. 

\item \emph{Entanglement}. Quantum entanglement is the principle that even if we know everything about the whole system, you don't necessarily know everything about its parts. The composite quantum system $AB$ could be in what we call a pure state, meaning that we have all of the information that characterizes that state as completely as the laws of physics will allow. But if we observe just part $A$ by itself, or part $B$ by itself, its state is not pure --- rather some of the information needed to completely characterize $A$ is missing, and likewise for $B$. Classical systems are not like that. In the classical world, if I know everything that can be known about the full system, I know everything about each of its parts as well. 
\end{description}

 \subsection{The qubit}
 The indivisible unit of classical information is the bit. We can encode a bit in any physical system that can be in either one of two possible distinguishable states. It is often convenient to think about bits abstractly, denoting the two possible states 0 and 1 without worrying about how the bit is physically instantiated. 
 
 The corresponding indivisible unit of quantum information is the \emph{qubit}. It too can be realized physically in many possible ways, but here we will mostly think about qubits abstractly, without any concern about the physical quantum system that encodes the qubit; it could be an atom, an electron, a photon, an electrical circuit, or something else.  While a bit is a two-state system that can be either 0 or 1, a qubit can be described mathematically as a vector in a complex Hilbert space, with two mutually orthogonal basis states which we can label $|0\rangle$ and $|1\rangle$:
 \begin{equation}
|\psi\rangle = a|0\rangle + b |1\rangle, \quad a,b\in \mathbb{C}, \quad |a|^2+|b|^2 = 1,\quad |\psi\rangle\sim e^{i\alpha}|\psi\rangle.
\end{equation}
It may seem like two complex numbers $a$ and $b$ are needed to describe the state of a qubit, but in fact just two real parameters suffice. That's because we don't really care about the overall normalization of the vector (by convention we usually set it equal to one). And we also don't care about its overall phase --- we can multiply the vector by a complex number of modulus one without changing anything that's physically relevant.  Once we have fixed the normalization, and removed the freedom to multiply by an overall phase, the general state of a qubit can be written as
\begin{equation}\label{psi-theta-varphi}
|\psi(\theta,\varphi)\rangle = e^{-i\varphi/2}\cos(\theta/2) |0\rangle + e^{i\varphi/2}\sin(\theta/2) |1\rangle, \quad 0\le\theta\le\pi,\quad 0\le \phi<2\pi.
\end{equation}

A bit is just the special case of a qubit where we know for sure the vector is always either $|0\rangle$ or $|1\rangle$. 
Imagine a game where Alice prepares a state of a qubit and sends the qubit to Bob; then Bob measures the qubit and tries to guess what state Alice sent. If Alice promises to always send either $|0\rangle$ or $|1\rangle$ then Bob can win the game every time. Orthogonal basis states can be perfectly distinguished by Bob if he does the appropriate measurement. 

The game becomes more challenging if the state sent by Alice is not necessarily chosen from among a set of mutually orthogonal states. In that case, there is no possible strategy for Bob that enables him to win the game for sure. Suppose for example that Alice promises to send to Bob one of the two states $|0\rangle$ or $|+\rangle$, where
\begin{equation}
|+\rangle = \frac{1}{\sqrt{2}}\left(|0\rangle + |1\rangle\right)\implies \quad \langle 0|+\rangle = \frac{1}{\sqrt{2}}.
\end{equation}
Now the possible states are not orthogonal, and no measurement by Bob can distinguish them perfectly. If the two states are equally likely, it turns out that if Bob plays the best possible strategy, he wins the game with probability $\cos^2(\pi/8)\approx .853$. Even though Bob knows for sure that Alice sent either $|0\rangle$ or $|+\rangle$, Bob has no way to find out for sure which of the two states he received. That's one important way a qubit is different from a bit.

More general measurements are possible, but we'll mostly be content to consider the simplest case of a measurement of a qubit, in which we force the qubit to become classical by declaring itself to be either $|0\rangle$ or $|1\rangle$. Then the rules of quantum theory say if the qubit state eq.(\ref{psi-theta-varphi}) is measured, the outcome will be $|0\rangle$ with probability $p_0=\cos^2(\theta/2)$ and will be $|1\rangle$ with probability $p_1=\sin^2(\theta/2)$. Even if Bob knows for sure the qubit is in the state $|\psi(\theta,\varphi)\rangle$, he can't predict the measurement outcome with certainty unless $\cos^2(\theta/2)$ is 0 or 1. That's another important way a qubit is different from a bit. 


\subsection{The tensor product}

To understand quantum computing we need to understand how we describe composite systems in quantum mechanics. Consider two individual quantum systems $A$ and $B$. How should I mathematically describe the composite system $AB$?

Let's say that system $A$ has dimension $d_A$. That means its Hilbert space $\mathcal{H}_A$ is spanned by a set of orthonormal basis states $\{|i\rangle_A, \, i=1, 2, \dots, d_A\}$. System $B$ has dimension $d_B$; it's Hilbert space $\mathcal{H}_B$ is spanned by the orthonormal basis states $\{|a\rangle, \, a = 1, 2, \dots, d_B\}$. Our rule for building the composite system should be compatible with the notion that if states are orthogonal then they can be perfectly distinguished.  That means we should define our inner product on the composite system this way:
\begin{equation}\label{tensor-basis}
\left(\langle j|\otimes \langle b|\right)\left(|i\rangle\otimes|a\rangle\right) = \delta_{ij}\delta_{ab}
\end{equation}
If I combine a basis state from system $A$ with a basis state from system $B$, then if the system $A$ basis states are mutually orthogonal, that means it's possible to perfectly distinguish the composite states. I can perfectly distinguish them just by looking at system $A$. Likewise, if I consider basis states for the composite system that are mutually orthogonal on system $B$, then those can be distinguished just by looking at system $B$. Reflecting this observation, the basis states in eq.(\ref{tensor-basis}) are orthogonal if either $i\ne j$ or $a\ne b$. What we have constructed is called the \emph{tensor product} $\mathcal{H}_A\otimes \mathcal{H}_B$ of the two Hilbert spaces.

For example, in the case of two qubits, we can choose our mutually orthogonal basis states to be 
\begin{equation}
\{|00\rangle,|01\rangle,|10\rangle,|11\rangle\},
\end{equation}
just as you would label the possible states of two bits. (To save writing, we sometimes use the notation $|00\rangle$ for the tensor product of vectors $|0\rangle\otimes |0\rangle$, etc.) These four vectors can be perfectly distinguished, and so should be regarded as mutually orthogonal vectors in the Hilbert space of the composite system. 

We can generalize this idea to many qubits. The Hilbert space for $n$ qubits is
\begin{equation}
\mathbb{C}^{2^n}= \underbrace{\mathbb{C}^2 \otimes\mathbb{C}^2\otimes \cdots\otimes\cdots  \otimes \mathbb{C}^2\otimes \mathbb{C}^2}_{n~\mathrm{times}}.
\end{equation}
This is the $2^n$-dimensional complex Hilbert space spanned by $2^n$ mutually orthogonal vectors labeled by bit strings:
\begin{equation}
|x\rangle = |x_{n-1}\rangle \otimes |x_{n-2}\rangle\otimes \cdots\otimes |x_1\rangle\otimes |x_0\rangle, \quad x\in\{0,1\}^n, 
\end{equation}
such that $\langle x|y\rangle = \delta_{x,y}$. If $x_i\ne y_i$ for the $i$th qubit, then I can distinguish the basis states just by observing the $i$th qubit, and correspondingly the basis states are orthogonal, no matter what the value of the other bits. 

The possible \emph{pure} quantum states of this $n$-qubit systems (those for which we have the most complete possible description) are vectors in that $2^n$-dimensional space with complex coefficients, 
\begin{equation}
|\psi\rangle=\sum_{x=0}^{2^n-1} a_x|x\rangle, \quad a_x\in \mathbb{C}.
\end{equation}
For a typical pure quantum state, specified as completely as the laws of physics will allow, a full description of that state in classical language requires a vast amount of classical data. For 300 qubits, say, the state is a vector in a space of dimension $2^{300}\approx 10^{90}$. We could not possibly write down a complete classical description of that state, even if every atom in the visible universe were at our disposal. This seems to be a fundamental principle about the extravagance of the quantum world: there is no succinct way to describe classically a typical quantum state of $n$ qubits. 

In the classical world, suppose I have a memory that can store $n$ bits. I might choose to store one of the possible strings of $n$ bits, and not tell you which string I stored. If you made a list of all the possible strings I might have chosen, that list would be unmanageably long. But in that classical setting, for any one of those $2^n$ possibilities, I can easily describe to you what I stored by sending to you just $n$ bits.
The quantum situation is radically different. For just one typical completely specified quantum state, there is no possible succinct classical description. 

There is an important caveat, though. While in the sense I just described a quantum state seems to be vastly more complex than a classical bit string, that very extravagant description of the quantum state is not directly accessible. If I send to you an $n$-qubit quantum state, and don't tell you anything about the state, you can try to learn something about the state by measuring it. Measuring forces the state to yield one particular $n$-bit string $x$ from among the $2^n$ possible strings. That is, the measurement yields only $n$ bits of classical data. You could try to extract classical information from my quantum state using some more clever strategy, but a rigorous result (\emph{Holevo's theorem} \cite{holevo1973bounds}) says there is no way to acquire more than $n$ bits of classical information from a single copy of an $n$-qubit quantum state, no matter what you do. The art of designing useful quantum algorithms hinges on somehow taking advantage of the hidden extravagance of the quantum world, overcoming our inability to glimpse any more than a meager shadow of the underlying quantum reality whenever we read out our quantum device.

When we speak of building a quantum system out of many qubits, one could ask: Given a Hilbert space of exponentially large dimension, how should I decompose it into qubits (or other lower-dimensional systems)? From a mathematical viewpoint the choice of a decomposition is completely arbitrary. But from a physical viewpoint there are preferred ways to do the decomposition. Typically, the way we choose to decompose a large quantum system as a tensor product of small systems is dictated by spatial locality. That is, we consider the different qubits to be at different locations. They might be in different cities, in Pasadena and New York, say. When we consider a quantum system shared between Pasadena and New York, it is very natural to decompose it in terms of qubits in Pasadena and qubits in New York. If I have $n$ atoms, which are spatially separated from one another, my preference would be to describe the system in terms of qubits residing in the individual atoms. 

The reason for that preference is that interactions between qubits are typically local in space. We expect that only systems in close proximity to one another interact significantly. The structure of these interactions provides a natural way to divide the big quantum systems into small subsystems. If qubits are located in $n$ different cities $A_1$, $A_2$, \dots $A_n$, it is quite easy to prepare a so-called product state
\begin{equation}
|\psi\rangle = |\psi_1\rangle_{A_1}\otimes |\psi_2\rangle_{A_2} \otimes \cdots \otimes |\psi_n\rangle_{A_n}.
\end{equation}
I can just call my friends in the $n$ cities and ask each one to prepare a single-qubit state. As we have seen, each of the single-qubit states is described by 2 real parameters, so any product state can be described succinctly by $2n$  real parameters. 

States that are not product states are called \emph{entangled} states. These are the states that cannot be created locally, no matter what I ask my $n$ friends to do in each city. Entangled states can be created only by quantum communication (sending qubits from one city to another) or by interactions among the qubits. My $n$ friends can talk on the (classical) telephone all they want, but if they don't have entanglement to begin with they cannot create entanglement --- they are stuck with a product state. To create entanglement shared among $n$ cities, we must either allow the qubits to come together to interact, or we could create an entangled state in one city, and then send each of those entangled qubits to different cities. These days, technological limitations make it hard to send a qubit a long way (say from Pasadena to New York) without the state of the qubit being damaged as it travels. But eventually we should be able to do that, allowing us to share a many-qubit entangled quantum state among the nodes of a quantum network distributed around the world. 

As we've seen, a typical entangled state of $n$ qubits is described by a number of classical parameters that is exponential in $n$, not linear in $n$ as for product states. But it turns out that if I want to make an entangled state of $n$ qubits, in principle I can create any such state, starting with a product state, if the qubits come together to interact two at a time. Mathematically, this is not a very deep fact. But physically it is quite significant. It means that by allowing particles to interact just two at a time, we can in principle control the quantum world.

There is a catch, though; we can't in general make any quantum state we desire efficiently, because the space of possible $n$-qubit states is so immense. The $n$-qubit Hilbert space has a dimension which is exponential in $n$, and a volume which is exponential in the dimension, hence \emph{doubly exponential} in $n$. If we consider all the ways I can bring together qubits to interact pairwise $T$ times in succession, roughly speaking the number of quantum states I can closely approximate is exponential in $T$. That means we need $T$ to be exponential in $n$ to create a state that closely approaches some specified typical $n$-qubit state. In this operational sense, the vast Hilbert space of $n$ qubits is to some degree illusory.  Future quantum engineers will know the joy of exploring Hilbert space far more extensively than we can today, but no matter how powerful their technology, most quantum states will remain far beyond their grasp. It's humbling to contemplate. 

\section{What is a quantum computer?}
\label{sec:computer}
 
 Now we know enough about quantum information to formulate a mathematical model of a quantum computer. 
 
 \subsection{The quantum circuit model}
\emph{Hilbert space}. To begin, we specify the arena in which a quantum computation takes place, namely the Hilbert space $\mathcal{H}=\mathbb{C}^{2^n}$. And it is important that the Hilbert space is not just a very large vector space, but is also equipped with a natural decomposition of the big Hilbert space as a tensor product of small subsystems, namely $n$ qubits. As already emphasized, this natural decomposition is in practice dictated by spatial locality. 
 
The reason it is so important to have a preferred decomposition is that we would like to be able to speak of the complexity of quantum computations or of quantum states. For a quantum state, I may define the complexity of the state as the number of steps need to prepare the state, but that number of steps depends on what the starting point is, and on what kind of operations we are allowed to perform in each step. Each elementary step should be something that's relatively easy to do in the lab, and we'll be guided by the idea that, while operations that act collectively on many qubits may be ``hard,'' operations that act on a small number of qubits (like one or two) are ``easy.''

\emph{Initial state}. The natural starting point, the initial state in a computation, is a product state. When we say ``product state'' we already have in mind a preferred decomposition into qubits. We may by convention say that each one of $n$ qubits starts out in the state $|0\rangle$. We consider starting with a product state because it is easy to prepare; we don't want to hide complexity in the initial state preparation itself.  The preparation is easy because I can set each qubit  to $|0\rangle$ separately, using operations that act on only one qubit at a time.  For example, I might measure each qubit in the basis $\{|0\rangle, |1\rangle\}$ (see below), and then flip the qubit if necessary to obtain the state $|0\rangle$.
 
\emph{Universal quantum gates.} Now, we consider building up a quantum state, or performing a computation, starting with the state $|0\rangle^{\otimes n}$. For that purpose, we'll need a specified set of elementary operations, which we can compose together in a circuit. Here, too, we need to distinguish between operations that are ``easy'' and operations that are ``hard,'' and again we make use of the preferred decomposition into qubits to make that distinction. The operations that act on a small number of qubits (a constant number, independent of $n$) are regarded as ``easy'', while generic operations that act on many qubits (a number that increases with $n$) are considered to be ``hard.'' We don't want to hide complexity in our choice of elementary operations. That way, it makes sense to quantify the complexity of a computation according to the number of elementary operations needed to execute the computation. 

Specifically, we suppose that there is a finite alphabet 
\begin{equation}
\mathcal{G} = \{U_1, U_2, \dots, U_{n_G} \}
\end{equation}
of unitary transformations, each acting on a constant number of qubits, which are hardwired in our quantum processor. A complex $m\times m$ matrix $U$ is unitary if $U^\dagger U= I$, where $U^\dagger$ denotes the adjoint of $U$. We consider unitary transformations because these are the ones that are allowed under the rules of how a finite-dimensional quantum system can evolve. (More general transformations are allowed if we pad our set of qubits with extra qubits, perform a joint unitary transformation on our qubits plus the extra ones, and then discard the extra ones. But in that case we could just consider those extra qubits to be included in the quantum computer's Hilbert space, so there is no loss of generality if we stick with only unitary operations.) We call these hardwired elementary unitaries our \emph{quantum gates}, as these are the quantum computer's counterpart of the elementary Boolean gates in the classical circuit model of computation. (Just to save syllables, we'll sometimes say ``unitaries'' as a shorthand for ``unitary transformations'' when the context makes the meaning clear.)
 
As previously mentioned, quantum gates acting on just two qubits are already universal --- with a circuit of two-qubit gates we can approximate any $n$-qubit unitary transformation as accurately as we please. Since two-qubit gates are usually easier to do in the physics lab than $k$-qubit gates for $k>2$, we normally include in our alphabet only gates acting on one or two qubits at a time.
 
 Now, unitary transformations, unlike Boolean classical logic gates, form a continuum of possible operations. An experimentalist who executes the gates has some classical parameters she can adjust to perform the gates; these parameters are real numbers, and as they vary continuously, so does the executed unitary transformation. So it may seem perverse to insist that our alphabet of quantum gates is a finite set, but there is actually a very good reason to do so. Since quantum hardware is not very reliable, we need to use quantum error-correcting codes to make our quantum gates more robust. And once we choose our code, only a finite set of one-qubit and two-qubit gates which are compatible with the structure of our code can be done efficiently and accurately. Only the robust gates from this finite set are good candidates to include in our gate set, and we should be sure these are chosen so as to maintain universality.

 If we want to accurately approximate any unitary transformation acting on a single qubit, it suffices to build circuits (sequences of gates) from just two noncommuting elementary gates. One popular choice that works (popular because these gates arise naturally when we use quantum error-correcting codes with nice properties) is
 \begin{equation}
 H = \frac{1}{\sqrt{2}}\left(
 \begin{array}{cc}
 1 & 1\\ 1 & -1
 \end{array}
\right),
 \quad
 T = \left(\begin{array}{cc}
 1 & 0\\ 0 & e^{i \pi/4}
 \end{array}
\right);
 \end{equation}
 $H$ is often called the ``Hadamard gate,'' and $T$ (for lack of a better name) is simply called the ``$T$ gate.''
 Since these two gates don't commute, sequences of gates chosen from this pair can reach a number of single-qubit unitaries that grows exponentially with the length of the sequence, and these densely fill the unitary group as the length increases. What is less obvious, but true and important, is that there is an efficient classical algorithm for finding a gate sequence that approximates a given desired one-qubit unitary within a specified error \cite{kitaev1997quantum,kitaev2002classical}. 
 
 With single-qubit gates and an initial product state we can reach only product states. But augmenting these single-qubit gates with just one entangling two-qubit gate is enough to achieve universality. A standard choice for this two-qubit gate (again, because it is convenient to perform on quantum information protected by a quantum error-correcting code) is the controlled-NOT (CNOT) gate with action
\begin{equation}
\mathrm{CNOT} = |0\rangle\langle 0|\otimes I+ |1\rangle\langle 1|\otimes X, \quad X = \left(
\begin{array}{cc} 0 & 1\\ 1 & 0\end{array}
\right).
\end{equation}
That is, if the first (control) qubit is $|0\rangle$, the CNOT gate acts trivially on the second (target) qubit, but if the control qubit is $|1\rangle$, the gate applies a bit flip ($X$) operation to the target qubit. The CNOT is a classical operation, in the sense that it maps our standard basis states to other standard basis states, but it can create entanglement when it acts on a control qubit that is in a superposition of basis states, for example:
\begin{equation}
\mathrm{CNOT}: \frac{1}{\sqrt{2}}\left(|0\rangle + |1\rangle\right)\otimes |0\rangle\rightarrow \frac{1}{\sqrt{2}}\left(|00\rangle + |11\rangle\right).
\end{equation}

Once we have fixed our universal gate set, we have a notion of how hard it is to reach a particular $n$-qubit unitary transformation, just as the circuit model of classical computation provides a notion of how hard it is to compute a particular Boolean function. 
We ask: what is the size of the minimal circuit that generates the desired unitary? A difference from the classical case is that, since the unitary transformations form a continuum, we in general have to accept some small error --- it may be that no circuit reaches the desired unitary exactly, in which case we settle for constructing a unitary that is only distance $\varepsilon$ way from the desired unitary according to some appropriate notion of distance. 
 
 So far, we have considered just one possible choice of the gates in our universal set $\mathcal{G}=\{H, T, \mathrm{CNOT}\}$. We had a good motivation for this choice, but for reasons of your own you might prefer a different universal gate set. Maybe you are using a different kind of hardware than I am, and there are some operations that are easy for you but not as easy for me, and vice versa. Even though we have different gate sets, we can agree on which unitaries are easy to construct and which ones are hard. That's because I can efficiently simulate your universal gates using mine, and you can efficiently simulate my universal gates using yours, with only modest overhead. Since my gates are universal, and each of your universal gates acts on a constant number of qubits, I can approximate any one of your universal gates, with an error $\varepsilon$, using polylog$(1/\varepsilon)$ of my gates. (That is, the number of gates I use scales like a polynomial in $\log(1/\epsilon)$ for asymptotically small $\epsilon$.) This fundamental result is called the \emph{Solovay-Kitaev theorem} \cite{kitaev1997quantum,kitaev2002classical}. 
 
 Let's say you can reach some $n$-qubit unitary $U$, with error $\delta$, using $T$ of your gates. Suppose I simulate each of your gates with error $\delta/T$, which I can do with $O(\mathrm{polylog} (T/\delta))$ of my gates. Since the error accumulates at worst linearly with the number of gates, I can simulate your circuit using $O\left(T (\mathrm{polylog} (T/\delta)\right)$ of my gates, making an error which is at worst $2\delta$. If your circuit is ``efficient'' (which we usually take to mean that $T$ is bounded above by some polynomial in the number of qubits $n$), then so is mine. 
 
 \emph{Classical control}. We do not want to hide complexity in the \emph{design} of our quantum circuit. The same issue arises in the circuit model of classical computation as well. In the classical case, we have to augment the circuit model with an additional computer, such as a Turing machine, which designs the circuit when we choose the problem we want to solve and specify the size of the input to the problem. In that case we demand that the runtime of this additional classical computer is also polynomial in the size of the input. Since a polynomial-size quantum circuit has a succinct classical description (even though the way the circuit acts on an input quantum state does not), we can use the same idea to augment the quantum circuit model. As in the classical circuit model, we insist that the family of quantum circuits that solve a problem with variable input size has the property of being \emph{uniform}. Informally, this just means that once we have found the circuit of size poly($n$) that works for the problem instance of specified size, the problem of finding the appropriate circuit for a larger input size is not much harder. 
 
 \emph{Readout}. So far we have been talking mostly about the complexity of a unitary transformation acting on a specified initial state. But when we use a quantum computer to solve a problem, we want the output of the computer to be classical information, which we can write down and share with our friends. To obtain a classical result, we need to do a measurement at the end of our quantum computation, and we don't want to hide complexity in our choice of readout procedure. Therefore, let's assume that we do the final readout by measuring the qubits (or some subset of the qubits) in the standard basis, obtaining the outcome 0 or 1 for each measured qubit. That completes our description of the quantum circuit model of computation. The initial state preparation and the final measurement are easy to do; what determines whether a quantum computation is easy or hard is the number of gate operations we need to do between the initial preparation and the final readout. 
 
I should emphasize that quantum computing is a randomized model of computation, because the measurement of a quantum state is not deterministic. If for example we are trying to solve a decision problem, for which the answer is either YES or NO, we might not get the correct answer every time we run the quantum computation. That's not really an issue, as long as we get the right answer with a sufficiently high success probability. The standard convention is to demand that our final readout yields the correct answer with probability at least $2/3$. Then if we run the computation a modest number of times, and take a majority vote (concluding that the answer is really YES if most of the computations gave the answer YES, and that the answer is NO if most of the computations gave the answer NO), then we will solve the problem correctly with probability close to one. 

\subsection{Computability and efficiency}
Now that we have formulated our model of quantum computation, we want to understand the power of the model. What computations can it run? What problems can it solve? In particular, what quantum algorithms can we design that achieve speedups relative to the best classical algorithms we know that solve the same problems?

To summarize, the features of the quantum model are these.
\begin{enumerate}
\item Scalable number of qubits.
\item Preparation of standard initial state.
\item Universal set of quantum gates.
\item Classical computer to design uniform quantum circuit families.
\item Readout in the standard basis.
\end{enumerate}
 
 \noindent We should emphasize that every feature of this model can be simulated by an ordinary classical computer, if equipped with a random number generator to capture the nondeterministic nature of the final quantum measurement. All the classical computer needs to do is keep track of a vector in a Hilbert space as we act on the vector with a sequence of matrices. For the final readout, we project the vector onto our standard set of axes, and assign probabilities to the different measurement outcomes accordingly. Since a (randomized) classical computer can do whatever a quantum computer does, there is no difference in \emph{computability} --- whatever is computable by a quantum computer is also computable by a classical computer. 
 
 The important distinction between the quantum and classical models is all about \emph{efficiency}. In general, for the classical computer to simulate the quantum computer, it has to deal with vectors in a space whose dimension is exponential in the number of qubits. For the hardest problem instances, we just don't know any way to do that simulation on the classical computer without using resources that scale exponentially with the number of qubits.
 
 From the viewpoint of physics (or the foundations of computer science), we should ask whether our abstract model of quantum computation is a good one for capturing the information processing that can really be done efficiently in the natural physical world. We don't know for sure whether that is the case or not. It's a hypothesis, what we might call the \emph{extended quantum Church-Turing thesis}. Whether this thesis is correct is not entirely obvious. To describe elementary particles the physicists use what they call local quantum field theory. Formally, quantum field theory involves a number of degrees of freedom per unit spatial volume which is infinite. We can't expect to simulate exactly an infinite number of degrees of freedom using any finite machine. It is an article of faith among physicists that we never really need all those degrees of freedom, that a good approximation can be attained by retaining only a finite number of degrees of freedom per unit volume. The number we need is controlled by input parameters of the simulation, such as the total energy of the process we wish to study. With a limited amount of energy, we cannot probe physics at arbitrarily short distance scales, so we only need a limited number of degrees of freedom to describe things accurately. In that case, we can argue persuasively that a physical process described by quantum field theory can be accurately and efficiently simulated using the quantum circuit model \cite{jordan2012quantum,preskill2018simulating}. 
 
Physicists believe that most phenomena in the physical universe can be accurately described by local quantum field theory. Possible exceptions arise, though, in situations where both quantum physics and gravitational physics play an important role. For example, we have only an incomplete understanding at present of how black holes process quantum information. 
Whether the quantum circuit model suffices for efficiently describing how black holes behave is something we don't yet know for sure. 
If the quantum circuit model really does capture everything that happens in physics, then we'll be able to use quantum computers in the future to  explore fundamental physics at a very deep level. But if not, that's even more exciting. It means that Nature will ultimately allow even more powerful information processors than the quantum computers we currently foresee.

 \subsection{Quantum hardware}
 
 The details of quantum hardware are beyond the scope of this article, but let's pause for a few words about how physical systems that we can really build and operate today align with the abstract model of quantum computing we have formulated. 
Though actual qubits are never perfect, we desire qubits that closely resemble the qubits described by the ideal model. Let's revisit the criteria our quantum hardware should satisfy, this time with an eye on whether actual devices are up to the task \cite{divincenzo2000physical}. 
 
 \begin{enumerate}
 \item Scalable system with controllable qubits.
\item Sufficiently accurate qubit preparation in the state $|0\rangle$.
\item Qubit coherence time sufficiently long compared to gate execution times. 
\item Sufficiently accurate universal set of quantum gates.
\item Sufficiently accurate qubit measurement in the standard basis.
 \end{enumerate}
 \noindent 
Various quantum systems have the potential to meet these desiderata; I'll just mention two. I apologize for the paucity of references in this discussion, but you can find more details in two recent reviews \cite{bruzewicz2019trapped,kjaergaard2020superconducting}.

 When Shor's algorithm precipitated a surge in interest in quantum computing in the mid-90s, it was a happy coincidence that experimental tools relevant to quantum computing were already being developed for other reasons. For one, advances in the technology for manipulating individual atomic ions trapped by electromagnetic fields had been motivated by the desire for more accurate atomic clocks. 
 
 In an ion trap, a single electrically charged atom (i.e. an ion), which can be in either its ground (lowest energy) state or some long-lived excited state, may serve as a qubit, and tens of such qubits may be loaded into a trap while maintaining precise control of each qubit. If we choose the right ion and the right excited state, then the other criteria can be met; in particular, idle qubits have very low error rates, i.e. the coherence time is very long, longer than a second for some ions. Furthermore, information processing, state preparation, and measurement can all be achieved by addressing the ions with pulses of light from a very stable laser. 
 
 For readout, one illuminates an atom with light of an appropriate frequency so that atoms in the ground state strongly scatter the light, while atoms in the excited state are transparent. Just by observing whether the illuminated ion glows or not, we can determine with high confidence whether the state of the qubit is $|0\rangle$ or $|1\rangle$. Measurement error rates below $10^{-4}$ can be achieved by interrogating an ion for a few hundred microseconds. Initial state preparation can also be achieved efficiently and accurately via laser manipulation of the ion's internal atomic state. 
 
Single-qubit quantum gates in ion traps are also easy and quite accurate. A laser pulse induces a coherent coupling between the two basis states of the qubit for a prescribed time to apply a desired unitary transformation to the qubit. Single-qubit gates with error rates below $10^{-4}$ can be executed in a few microseconds. 
 
For ion traps, as for most other quantum platforms, the most challenging task is performing entangling two-qubit gates, which require that two atoms interact sufficiently strongly. The electrostatic repulsion of the ions provides the needed interaction. Because of the repulsion, the ions have shared normal modes of vibration in the trap. A laser pulse couples a normal mode shared by two ions to the internal state of the pair of ions, guiding that mode on an excursion during which the two-qubit state acquires a phase that depends on the internal states of the two atoms; the result is an entangling two-qubit gate \cite{molmer1999multiparticle,sorensen1999quantum}. The speed of the gate depends on the optical power of the laser and the vibrational frequencies of the ions in the trap; it typically takes at least tens of microseconds. Gates are usually executed sequentially rather than in parallel, to avoid unwanted couplings between qubits that might compromise gate fidelity. In the best current multi-qubit devices, the error rate per entangling two-qubit gate is typically around 1\%, though error rates below $10^{-3}$ have been achieved under highly favorable conditions. 

As an alternative to actual atoms, engineered ``artificial atoms'' may serve as qubits. In particular, reasonably high-quality qubits can be realized using superconducting electrical circuits, which conduct electricity with negligible resistance at sufficiently low temperature. These circuits have an energy-level structure reminiscent of an atom's if the circuit includes nonlinear elements (Josephson junctions), and a qubit can be encoded using the circuit's lowest energy state $|0\rangle$ and its first excited state $|1\rangle$. The energy splitting between these levels is typically around 5 GHz, and the device is kept at a temperature (10--20 mK $\approx$ 200--400 MHz) which is sufficiently small compared to this splitting that the thermal state of the qubit is very close to its quantum-mechanical ground state. Coherence times of tens to hundreds of microseconds can be routinely achieved. 

The scheme for executing single-qubit gates is conceptually comparable to the scheme used in ion traps, except that the qubit's evolution is driven by a microwave pulse rather than a laser. If the pulses are carefully shaped to avoid unwanted transitions to higher energy levels, single-qubit gate error rates well below 1\% can be achieved in a few tens of nanoseconds. 

There are several different schemes for performing entangling two-qubit gates. For example, one can tune the frequency of a qubit by applying a magnetic flux, and the desired gate can be obtained by bringing two quantum states of a pair of qubits to nearly coincident frequencies for a specified time. In multi-qubit devices, two-qubit gate error rates comparable to 1\% can be achieved in tens of nanoseconds. (As is the case for ions, two-qubit gate error rates below $10^{-3}$ have now been achieved under highly favorable conditions.)

To read out a qubit, one couples it to a microwave resonator, and the resonator's frequency shifts by an amount that depends on whether the qubit's state is $|0\rangle$ or $|1\rangle$. The frequency shift can then be detected by observing the resonator's response to a microwave drive. A measurement error rate of about 1\% can be achieved in hundreds of nanoseconds. 

Ion traps and superconducting circuits are currently the two leading qubit technologies, and each has characteristic advantages and disadvantages. For example, atoms are all alike, and have exceptionally long coherence times. In addition, a two-qubit gate can be performed acting on any pair of ions in a trap, with a fidelity that is not very sensitive to the spatial separation between the ions.
In contrast, the coherence times of superconducting qubits are limited by imperfections in how they are fabricated; furthermore their properties vary from qubit to qubit and can evolve in time as well. Therefore the qubits need to be carefully and frequently calibrated. Also, though schemes for long-range coupling have been proposed, in today's state-of-the-art quantum processors high-quality two-qubit gates are performed only between neighboring qubits laid out in a one-dimensional or two-dimensional array.

On the other hand, quantum gates are much faster in superconducting devices, and many gates can be executed in parallel without the gate fidelity being seriously diminished. That may be a big advantage in the future, when we judge the performance of a quantum computation according to the total time needed to find a solution. 

Scaling from the tens of qubits we have now to the millions of physical qubits we expect to need in the future will be daunting challenge for both ion traps and superconducting circuits, just as for all the other currently known quantum technologies. With more than about 100 ions in a  trap, it becomes too difficult to control all of the coupled vibrational modes. To scale up further will probably require some sort of modular design, with many relatively small traps networked together into a large system. To share quantum information among the modules, several ideas have been suggested. We might engineer optical interconnects, which allow a photon to travel coherently from one trap to another. Or we might shuttle ions between traps while maintaining the coherence of the ions' internal states. Both these approaches are under development, but still have far to go before a large-scale modular quantum computer becomes practical. 

For superconducting circuits as well, control of the system becomes increasingly challenging as the number of qubits increases, in part because of the proliferation of microwave control lines which exacerbates problems like crosstalk. Some of these issues can be mitigated through advances in engineering, but there are big opportunities in basic research, too. Superconducting circuits can support a rich variety of device designs, and there are many possibilities yet to be explored. 

Many other promising approaches to quantum hardware are being pursued, but we won't delve into the subject any further here.  Above all, I want to emphasize that we are still in the early stages of developing quantum computing systems, and no doubt big surprises lie ahead. The brief synopsis above is likely to be badly out of date soon!

 
\section{Simulating quantum dynamics}
\label{sec:time-evolution}

Next we'll look in more depth at how quantum computers can be used to solve problems in quantum physics, as foreseen by Feynman. An especially important application is solving the time-dependent Schr\"{o}dinger equation, i.e. finding out how an $n$-quantum system evolves in time, as governed by some many-body Hamiltonian. In special cases we know how to solve this problem efficiently with a classical computer; sometimes we can even find an analytic solution. But in general the best classical algorithms have a runtime that scales exponentially with $n$. Simulating time evolution with a quantum computer, in contrast, scales polynomially with $n$ if the Hamiltonian $H$ is \emph{local} \cite{lloyd1996universal}. Here we'll explain why this exponential quantum speedup is possible, without attempting to exhibit the best state-of-the-art quantum algorithms. 

For a system of $n$ qubits, we say that $H$ is {\it k-local} if 
\begin{equation}
H = \sum_a H_a,
\end{equation}
where each term $H_a$ acts non-trivially on at most $k$ qubits --- i.e. $H_a = \tilde{H}_a \otimes I^{n-k}$, and $\tilde{H}_a$ acts on some set of at most $k$ qubits. (Of course, we may use a similar definition for a system of $d$-dimensional subsystems for constant $d > 2$, rather than qubits.) We say that $H$ is local if it is $k$-local for some constant $k$. 

There is a stronger notion of locality we sometimes use, which can be called {\it geometrical locality} or {\it spatial locality}. A $k$-local Hamiltonian is geometrically local in $D$ dimensions if the qubits can be arranged in (flat) $D$-dimensional space with a bounded number of qubits per unit volume, and the $k$ qubits upon which $H_a$ acts non-trivially are all contained in a ball of constant radius. In this sense there are no {\it long-range} interactions among the qubits. $H$ is geometrically local if it is geometrically $k$-local in $D$ dimensions for some constant $D$ and $k$.

If we write $H = \sum_a H_a$ where there is a unique $H_a$ for each set of $k$ qubits, then the expansion of a $k$-local $H$ contains at most $\binom{n}{k} = O(n^k)$ terms, and the expansion of a geometrically local $H$ contains $O(n)$ terms (each of the $n$ qubits is contained in a constant number of interacting sets). Let us also assume that each $H_a$ is bounded:
\begin{equation}
||H_a ||_\infty \leq h ~~\text{for all $a$, where $h$ is a constant}.
\end{equation} 
Physicists are interested in geometrically local Hamiltonians because they seem to provide an accurate description of Nature. Therefore, it is noteworthy that quantum circuits can simulate quantum evolution governed by a local Hamiltonian efficiently: evolution of $n$ qubits for time $t$ can be simulated to constant accuracy using a circuit whose size is polynomial in $n$ and $t$.

We can formulate the problem this way: suppose we are given an initial quantum state $|\psi(0) \rangle$, or a classical description of a quantum circuit that prepares the state. Our goal is to construct
\begin{equation}
|\psi(t) \rangle = U(t) | \psi(0) \rangle 
\end{equation}
where $U(t)$ satisfies $\frac{d}{dt}U(t) = -i H(t) U(t)$ and the boundary condition $U(0)=I$. (Thus $U(t)=e^{-iHt}$ in the case where $H$ is time independent.) We will settle for computing $|\psi(t) \rangle$ to accuracy $\delta$, i.e. constructing $\tilde{\psi}(t)\rangle$ where
\begin{equation}
|| |\tilde{\psi}(t)\rangle - |\psi(t)\rangle || < \delta.
\end{equation}
Depending on the situation, we might be satisfied if $\delta$ is a sufficiently small constant, or we might impose the stricter requirement that the error is smaller than some specified power of the size $n$ of the system. 
To relate this simulation task to a task that can be described classically, suppose the goal is to sample from the probability distribution
\begin{equation}
\langle \psi(t) | \Pi_a | \psi(t) \rangle
\end{equation}
where $\Pi_a$ projects onto an eigenstate with eigenvalue $a$ of an observable $A$ that can be measured efficiently by a quantum computer. Classically this task is believed to be hard at least in some cases, because the unitary matrix $U(t)$ is exponentially large $(2^n \times 2^n)$. But quantumly we can do the simulation efficiently if $H$ is a local Hamiltonian. 

To simulate continuous time evolution on a classical or quantum computer, we choose a small step size $\Delta$, and approximate evolution for time $t$ by a sequence of $t/\Delta$ steps. (If $H$ is actually time dependent, assume $\Delta$ is small enough that the change of $H$ during a time interval of width $\Delta$ can be neglected.) We wish to attain accuracy
\begin{equation}
|| \tilde{U}(t) - U(t) ||_\infty < \delta,
\end{equation}
where $\tilde{U}$ is the simulated unitary and $U$ is the ideal unitary. Hence the error per time step should be less than $\delta \Delta/t$. 

Suppose $H = \sum_a H_a$ is a sum of $M$ $k$-local terms, and let's consider the geometrically local case, where $M = O(n)$. We will show below that a single time step can be simulated by a product of $M$ local ``gates'' (unitary transformations that act on a constant number of qubits) where each such ``gate'' has an error $O(\Delta^2 h^2)$. Therefore the simulation of evolution for time $t$ uses all together $Mt/\Delta$ {gates} where we require
\begin{equation}
\frac{Mt}{\Delta} \Delta^2 h^2 \approx \delta ~~ \implies ~~ \Delta  = O \left(\frac{\delta}{h^2 M t} \right).
\end{equation}
Therefore the total number of gates is
\begin{equation}
L  = O \left(\frac{h^2(Mt)^2}{\delta} \right).
\end{equation}
Furthermore each ``{gate}'' can be simulated to accuracy $O(\Delta^2 h^2)$ with a universal gate set using ${\rm polylog} \left(\frac{1}{\Delta^2 h^2} \right) = {\rm polylog} \left(\frac{h^2(Mt)^2}{\delta^2} \right)$ gates, according to the Solovay-Kitaev theorem. We conclude that the simulation can be done with a quantum circuit of size 
\begin{equation}\label{eq:simulation gate count}
L  = O \left(\frac{h^2(Mt)^2}{\delta} {\rm polylog} \left( \frac{h^2(Mt)^2}{\delta^2} \right) \right).
\end{equation}
In the case where $H$ is  geometrically local, $M = O(n) = O(V)$, where $V$ is the spatial volume of the system. Since $h$ is a constant, we find  that the cost of simulating time evolution with fixed accuracy scales like 
\begin{equation}
L  = O(\Omega^2~{\rm polylog} ~\Omega),
\end{equation}
where $\Omega = Vt$ is the simulated volume of spacetime.

Now we need to explain how to simulate a single time step. We'll use the idea that $\exp{\left(\sum_a A_a\right)}$ can be approximated by $\prod_a \exp{( A_a)}$ if $||A|| \ll 1$. To check the accuracy we expand the exponentials:
\begin{eqnarray}
&&\exp{\left(\sum_a A_a\right)} - \prod_a \exp{\left( A_a\right)} \\
&=& \left( 1 + \sum_a A_a + \frac{1}{2} \sum_{a,b} A_a A_b +\dots \right) - \prod_a \left(  1 + A_a + \frac{1}{2}A_a^2 + \dots \right)  \nonumber \\
&=& \left( 1 + \sum_a A_a + \frac{1}{2} \sum_{a,b} A_a A_b +\dots\right) - \left(  1 + \sum_a A_a + \sum_a \frac{1}{2} A_a^2 + \sum_{a<b} A_a A_b + \dots\right)  \nonumber \\
&=& \frac{1}{2}\left( \sum_{a<b} A_a A_b + \sum_{a<b} A_b A_a \right) - \sum_{a<b} A_a A_b + \dots  \nonumber \\
&=& -\frac{1}{2} \sum_{a<b} [A_a, A_b]  + \dots  \nonumber
\end{eqnarray}
(where $+\dots $ denotes terms higher order in $A_a$). Writing $H = \sum_a H_a$, then, we find that 
\begin{equation}
e^{-i H \Delta} - \prod_a e^{-i H_a \Delta} = \frac{1}{2} \Delta^2 \sum_{a < b} [H_a,H_b] + {\rm higher ~order}.
\end{equation}

Now, how many non-vanishing commutators $\{ [H_a, H_b]\}$ can occur in this sum? 
Let's suppose the Hamiltonian is geometrically local, in which case there are $O(n)$ terms in $H$, and each term fails to commute with a constant number of terms. So, there are $O(n) = O(M)$ non-vanishing commutators. 
We conclude that (in the geometrically local case)
\begin{equation}
\left\| e^{-iH \Delta} - \prod_a e^{-iH_a \Delta} \right\| = O(M \Delta^2 h^2).
\end{equation}
Since $\Pi_a e^{-iH_a \Delta}$ is a product of $M$ ``gates,'' we have verified that the accuracy per gate is $O(\Delta^2 h^2)$. (Note that terms arising from the higher-order terms in the expansion of the exponential are of order $M \Delta^3 h^3$, and therefore systematically suppressed by another factor of $\Delta h = O(\delta/hMt) = O( ( \delta/L)^{1/2})$. )

We have shown that, for a geometrically local $H$ that is a sum of bounded terms, evolution in a spacetime volume $\Omega$ can be achieved with a quantum circuit of size
\begin{equation}
L = O(\Omega^2~ {\rm polylog}~ \Omega),
\end{equation}
The simulation can be achieved with quantum resources which scale like the {\it square} of the simulated volume (up to a polylog factor). With more sophisticated methods, the scaling with $\Omega$ and also the scaling with the error $\delta$ can be improved. We will not discuss these improvements here, even though they may be of great practical importance in the future if they can substantially reduce the runtime on a quantum computer for problems of interest to physicists and chemists. We have settled for making the crucial point --- that the quantum runtime scales polynomially with the size of the physical system, while the best general-purpose classical algorithms scale exponentially. 

\section{Energy eigenvalues and  eigenstates}
\label{sec:energy}

Aside from simulating time evolution, physicists and chemists are also interested in ``diagonalizing'' many-body Hamiltonians, i.e. finding energy eigenvalues and properties of energy eigenstates. Here again there are special cases where we can find analytic solutions or obtain good approximate solutions efficiently using classical computers. But there are many cases of physical interest where the problem seems to be hard classically, simply because the Hamiltonian is an extremely large $2^n\times 2^n$ matrix. With a quantum computer we can ``solve'' the problem efficiently, subject to some important caveats which we'll come to. 

The algorithm for estimating eigenvalues and preparing eigenstates of a local Hamiltonian $H$ using a quantum computer makes use of the algorithm described in Sec.~\ref{sec:time-evolution} for simulating time evolution. Once we have constructed an efficient quantum circuit for the time-evolution operator $U(t)=\exp(-i Ht)$, we apply a general procedure for estimating eigenvalues of unitary matrices. This general procedure, called \emph{phase estimation} \cite{kitaev1995quantum}, leverages a very efficient procedure for evaluating the Fourier transform on a quantum computer. Phase estimation is an essential primitive used in a variety of quantum algorithms, including Shor's factoring algorithm. 

\subsection{Quantum Fourier transform}

Before explaining phase estimation, let's see how to Fourier transform a function using a quantum computer. We suppose that the function is encoded in the \emph{amplitudes} of an $m$-qubit quantum state:
\begin{equation}
\sum_{x=0}^{N-1} f(x)|x\rangle;
\end{equation}
here $x=x_{m-1}x_{m-2}\dots x_1 x_0$ is shorthand for the integer $x$ expanded in binary notation, and $N=2^m$.
The discrete quantum Fourier transform (QFT) acts on this state according to
\begin{equation}
\mathrm{QFT}: \sum_{x=0}^{N-1} f(x) |x\rangle \rightarrow \sum_{k=0}^{N-1} \left({1\over\sqrt{N}} \sum_{x=0}^{N-1} e^{2\pi
i kx/N} f(x)\right)|k\rangle;
\end{equation}
it is an $N\times N$ unitary matrix with matrix elements $\{(e^{2\pi i/N})^{kx}/\sqrt{N}\}$. Here $N$ might be exponentially large, but thanks to the simple structure of the QFT, it can be implemented by a quite efficient quantum circuit containing only $O(m^2)$ gates. 

If we express $x$ and $k$ as binary expansions
\begin{align}
x &= x_{m-1} \cdot 2^{m-1} + x_{m-2} \cdot 2^{m-2} + \ldots + x_1 \cdot 2 +
x_0,\notag \\
k &= k_{m-1} \cdot 2^{m-1} + k_{m-2} \cdot 2^{m-2} + \ldots + k_1 \cdot 2 +
k_0,
\end{align}
then in the product of $x$ and $k$, we may discard any terms containing $m$ or more
powers of $2$, as these make no contribution to $\exp(2\pi i kx/2^m)$.  Hence
\begin{align}\label{bin_product}
{kx\over 2^m} &\equiv k_{m-1} (. x_0) + k_{m-2} (. x_1 x_0) + k_{m-3} (. x_2
x_1 x_0) + \ldots \notag \\
&+ k_1 (. x_{m-2} x_{m-3} \ldots x_0) + k_0 (. x_{n-1} x_{m-2} \ldots x_0),
\end{align}
where the factors in parentheses are binary expansions; e.g.,
\begin{equation}
. x_2 x_1 x_0 = {x_2\over 2} + {x_1\over 2^2} + {x_0\over 2^3}.
\end{equation}
Using eq.(\ref{bin_product}), we can see that the quantum Fourier transform maps each computational basis state to a product state of $m$ qubits:
\begin{align}
\mathrm{QFT}: &|x\rangle \rightarrow {1\over\sqrt{N}} \sum_{k=0}^{N-1} e^{2\pi i
kx/N}|k\rangle\notag \\
& = {1\over\sqrt{2^m}} \left(\underbrace{|0\rangle + e^{2 \pi i(. x_{0})} |1\rangle}_{k_{m-1}}\right)\otimes
\left(\underbrace{|0\rangle + e^{2 \pi i(. x_{1} x_{0})} |1\rangle}_{k_{m-2}}\right)\otimes \notag \\
& \ldots \otimes\left(\underbrace{|0\rangle + e^{2 \pi i(. x_{m-1} x_{m-2} \ldots x_{0})}
|1\rangle}_{k_0}\right);
\end{align}
as a result, it can be efficiently implemented.  To be concrete, 
consider the case $m = 3$.  We can readily see that the circuit
\begin{center}

\begin{picture}(290,100)

\put(0,74){\makebox(20,12){$\ket{x_2}$}}
\put(0,44){\makebox(20,12){$\ket{x_1}$}}
\put(0,14){\makebox(20,12){$\ket{x_0}$}}

\put(270,74){\makebox(20,12){$\ket{k_0}$}}
\put(270,44){\makebox(20,12){$\ket{k_1}$}}
\put(270,14){\makebox(20,12){$\ket{k_2}$}}

\put(20,80){\line(1,0){10}}
\put(50,80){\line(1,0){20}}
\put(90,80){\line(1,0){20}}
\put(130,80){\line(1,0){130}}

\put(20,50){\line(1,0){130}}
\put(170,50){\line(1,0){20}}
\put(210,50){\line(1,0){50}}

\put(20,20){\line(1,0){210}}
\put(250,20){\line(1,0){10}}

\put(80,50){\circle*{4}}
\put(120,20){\circle*{4}}
\put(200,20){\circle*{4}}

\put(80,70){\line(0,-1){20}}
\put(120,70){\line(0,-1){18}}
\put(120,48){\line(0,-1){28}}
\put(200,40){\line(0,-1){20}}

\put(30,70){\framebox(20,20){$H$}}
\put(70,70){\framebox(20,20){$R_1$}}
\put(110,70){\framebox(20,20){$R_2$}}

\put(150,40){\framebox(20,20){$H$}}
\put(190,40){\framebox(20,20){$R_1$}}

\put(230,10){\framebox(20,20){$H$}}

\end{picture}
\end{center}
\noindent does the job (but note that the order of the bits has been reversed
in the output).  Each Hadamard gate $H$ acts as
\begin{equation}
H: |x_j\rangle \rightarrow {1\over\sqrt{2}} \left(|0\rangle +(-1)^{x_j}|1\rangle\right)={1\over\sqrt{2}} \left(|0\rangle + e^{2\pi i(.
x_{j})} |1\rangle\right).
\end{equation}
The other contributions to the relative phase of $|0\rangle$ and $|1\rangle$ in
the $j$th qubit are provided by the two-qubit controlled rotations, where
\begin{equation}
R_d = \left(\begin{array}{ll} 1 & 0\\ 0 & e^{i\pi/2^{d}}\end{array}\right),
\end{equation}
and $d = (j-\ell)$ is the ``distance'' between the qubits. (The controlled $R_d$ shown in the circuit diagram applies the nontrivial phase $e^{i\pi/2^{d}}$ only if the two-qubit state is $|11\rangle$).

In the case $m = 3$, the QFT is constructed from three $H$ gates and three
controlled-$R_d$ gates.  For general $m$, the obvious generalization of this
circuit requires $m$ $H$ gates and $\left({m\atop 2}\right) = {1\over 2} m
(m-1)$ controlled $R_d$'s.  A two-qubit gate is applied to each pair of qubits,
again with controlled relative phase $\pi/2^d$, where $d$ is the ``distance''
between the qubits.  Thus the circuit family that implements the QFT has a size of
order $(\log N)^2$. On a quantum computer, the Fourier transform is remarkably easy to implement, even when $N$ is exponentially large. In contrast, the classical ``fast Fourier transform'' algorithm has a runtime $O(N\log N)$.

\subsection{Phase estimation}\label{subsec:phase-est}

Phase estimation is a quantum algorithm that  estimates eigenvalues of a unitary operator $U$, using the QFT as a subroutine.
 The quantum circuit makes use of an auxiliary register which records an integer-valued ``time'' parameter $t$; this time register is initialized in a uniform superposition of all values of $t$ running from $t=0$ to $t=2^m -1$. Then the unitary $U$ is executed $t$ times acting on a data register, controlled by the time register. If the initial state of the data register is $|\psi\rangle$, this procedure prepares the state
 \begin{equation}
 \frac{1}{\sqrt{2^m}}\left(\sum_{t=0}^{2^m-1} |t\rangle\otimes U^t |\psi\rangle\right).
 \end{equation}
 
 To be more concrete, the circuit that prepares this state is shown here for the case $m=3$:

\begin{center}

\begin{picture}(300,130)

\put(0,104){\makebox(20,12){$\ket{0}$}}
\put(0,74){\makebox(20,12){$\ket{0}$}}
\put(0,44){\makebox(20,12){$\ket{0}$}}
\put(0,14){\makebox(20,12){$\ket{\lambda}$}}

\put(210,104){\makebox(60,12){${1\over\sqrt{2}}\left(\ket{0}+\lambda^4\ket{1}\right)$}}
\put(210,74){\makebox(60,12){${1\over\sqrt{2}}\left(\ket{0}+\lambda^2\ket{1}\right)$}}
\put(210,44){\makebox(60,12){${1\over\sqrt{2}}\left(\ket{0}+\lambda\ket{1}\right)$}}

\put(20,110){\line(1,0){10}}
\put(50,110){\line(1,0){140}}

\put(20,80){\line(1,0){10}}
\put(50,80){\line(1,0){140}}

\put(20,50){\line(1,0){10}}
\put(50,50){\line(1,0){140}}

\put(20,25){\line(1,0){50}}
\put(20,20){\line(1,0){50}}
\put(20,15){\line(1,0){50}}

\put(90,25){\line(1,0){20}}
\put(90,20){\line(1,0){20}}
\put(90,15){\line(1,0){20}}

\put(130,25){\line(1,0){20}}
\put(130,20){\line(1,0){20}}
\put(130,15){\line(1,0){20}}

\put(170,25){\line(1,0){20}}
\put(170,20){\line(1,0){20}}
\put(170,15){\line(1,0){20}}

\put(80,50){\circle*{4}}
\put(120,80){\circle*{4}}
\put(160,110){\circle*{4}}

\put(80,50){\line(0,-1){20}}
\put(120,80){\line(0,-1){28}}
\put(120,48){\line(0,-1){18}}
\put(160,110){\line(0,-1){28}}
\put(160,78){\line(0,-1){26}}
\put(160,48){\line(0,-1){18}}
\put(120,80){\line(0,-1){28}}
\put(120,80){\line(0,-1){28}}

\put(30,100){\framebox(20,20){$H$}}
\put(30,70){\framebox(20,20){$H$}}
\put(30,40){\framebox(20,20){$H$}}
\put(70,10){\framebox(20,20){$U$}}
\put(110,10){\framebox(20,20){$U^2$}}
\put(150,10){\framebox(20,20){$U^4$}}

\end{picture}
\end{center}
\noindent The three Hadamard gates acting on $|0\rangle^{\otimes 3}$ prepare the uniform superposition of $2^3$ computational basis  states $\{|t_2 t_1 t_0\rangle\}$. Then $U$ is applied conditioned on the least significant bit $|t_0\rangle$, $U^2$ conditioned on the next bit $|t_1\rangle$, and so on. If the initial state of the data register happens to be an eigenstate $|\lambda\rangle$ of $U$ with eigenvalue $\lambda$, this circuit yields the state
\begin{equation}
\frac{1}{\sqrt{2^3}}\left(\underbrace{\ket{0}+\lambda^4\ket{1}}_{t_2}\right)\otimes \left(\underbrace{\ket{0}+\lambda^2\ket{1}}_{t_1}\right)\otimes \left(\underbrace{\ket{0}+\lambda\ket{1}\rangle}_{t_0}\right) = \frac{1}{\sqrt{2^3}}\sum_{t=0}^7\lambda^t |t\rangle\otimes |\lambda\rangle. 
\end{equation}

To recover the value of $\lambda$, we can now apply the QFT to the time register and measure in the computational basis. If $\lambda = e^{-2\pi i k/ 2^m}$, where $k= k_{m-1}k_{m-2}\dots k_1k_0$ is an integer less than $2^m$, then the measurement outcome will be $k$ with probability one. More generally, if $\lambda=e^{-2\pi i \phi}$ where the binary expansion of $\phi$ does not necessarily terminate after $m$ bits, the measurement finds $\phi$ to about $m$ bits of precision with high success probability. In other words, we can estimate $\phi$ with an accuracy $\delta\approx 2^{-m}$ by conditionally applying $U$ up to $2^m\approx 1/\delta $ times. 

If the initial state of the data register is not an eigenstate of $U$, it can be expanded in terms of $U$ eigenstates. If we apply the phase estimation circuit and obtain the measurement outcome $k$, then the data is projected onto the $U$ eigenstates with eigenvalues that are close to $e^{-2\pi i k/ 2^m}$. Once an (approximate) eigenstate of $U$ has been prepared in this fashion, we can perform additional measurements to collect further information about the properties of this state. Since both this preparation of the $U$ eigenstate and the additional measurements we perform on that eigenstate are nondeterministic, we may need to repeat the whole procedure multiple times to acquire statistically useful information. Furthermore, the probability of finding a particular eigenstate will of course depend on the initial state of the data register to which phase estimation is applied.

\subsection{Hamiltonian eigenstates}

If we can simulate quantum evolution governed by a Hamiltonian $H$, then we can use the phase estimation algorithm to find eigenvalues and prepare eigenstates of $H$. To obtain eigenvalues to $m$ bits of accuracy, we choose a convenient unit of time $T$, and execute the time evolution operator $e^{-i H s}= U^t$, where $U=e^{-iHT}$, conditioned on 
$t \in  \{ 1,2, 4, 8, .. 2^{m-1} \}$. That is, the control parameter $t$ used in phase estimation may now be interpreted as the evolution time $s$ expressed in units of $T$. Note that if there is an efficient circuit for $U^t$, then $U^t$ conditioned on a single control qubit is also efficient, with a comparable gate count.

\begin{figure}
\begin{center}
\includegraphics[width=\textwidth]{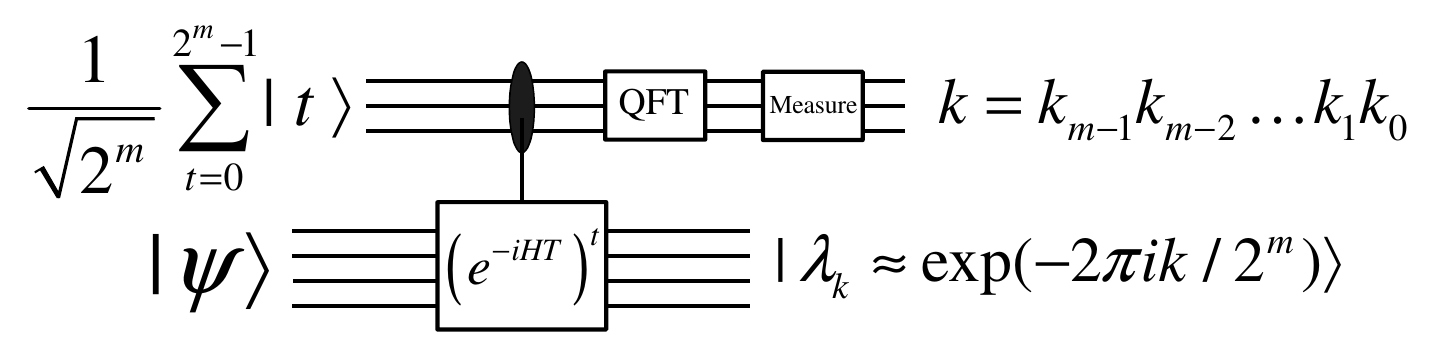}
\end{center}
\caption{\label{fig:phase-estimation} Phase estimation algorithm for measuring eigenvalues of $e^{-iHT}$.}
\end{figure}

As in Sec.~\ref{subsec:phase-est}, phase estimation then suffices to find the fractional part of $\frac{ET}{2\pi}$ to $m$-bit accuracy, where $E$ is an eigenvalue of the Hamiltonian $H$. We should choose the step size in the simulation of $e^{-iHs}$ so that the accuracy is $\delta \approx 2^{-m}$ for $s= 2^m T$. If the Hamiltonian is geometrically local, we have seen in eq.(\ref{eq:simulation gate count}) 
that this approximation can be achieved with a circuit size 
\begin{equation}
L=\tilde O\left(\frac{h^2(ns)^2}{\delta}\right)=\tilde O\left(h^2(nT)^2\times \frac{2^{2m}}{2^{-m}}\right)=\tilde O\left((hT)^2n^22^{3m}\right).
\end{equation}
(Here we use the $\tilde O$ notation to indicate that a polylog factor has been neglected.)
To compute the energy eigenvalue to accuracy polynomial in the system size $n$, we choose 
\begin{equation}
\delta\approx 2^{-m} \approx 1/n^c \implies m = c ~\log_2 n,
\end{equation} 
where $c$ is a constant.
The algorithm is efficient ---  the quantum circuit size is
\begin{equation}
\tilde O \left((hT)^2 n^2 2^{3m}\right) = \tilde O (n^2n^{3c}),
\end{equation} 
which is polynomial in $n$. The approximations we have used can be improved significantly; my goal here was just to explain as simply as possible why there is an exponential quantum advantage. The phase estimation algorithm for measuring $e^{-iHT}$ is shown schematically in Fig.~\ref{fig:phase-estimation}.

For a particular preparation of the input state $|\psi \rangle$, suppose we repeat the computation many times, and plot a histogram of the results. Then the {\it location} of each narrow peak estimates an energy eigenvalue $E_a$, modulo $2\pi/T$. The {\it height} of the peak estimates $| \langle E_a | \psi \rangle |^2$ -- the overlap $|\psi \rangle$ with the corresponding energy eigenstate $|E_a\rangle$. 

However, if we want to estimate (say) the ground state energy $E_0$ to polynomial accuracy in quantum polynomial time, we must be able to prepare a state $| \psi \rangle$ whose overlap with the ground state $|E_0 \rangle$ is no worse than polynomially small:
\begin{equation}
|\langle E_0 | \psi \rangle |^2 > 1/{\rm poly}(n).
\end{equation} 
If that is the case, we can get a good estimate of $E_0$ in only polynomially many trials. As a bonus, when we obtain the value $E_0$ for the measured eigenvalue $E_0$, then we have projected the state $|\psi \rangle$ onto the ground state $|E_0 \rangle$, and therefore we can compute further properties of $|E_0 \rangle$, such as the distribution ${\rm Prob}(a) = \langle E_0 | \Pi_a | E_0 \rangle$, where $\Pi_a$ is a projector onto an eigenspace of an efficiently measurable observable.

\subsection{Initial state preparation}

However, there is a catch --- preparing an initial state that overlaps substantially with the ground state could be very hard in some cases. This is already true classically; finding a good approximation to the ground state of a classical spin glass is NP-hard, as hard as any problem whose solution can be checked efficiently by a classical computer \cite{barahona1982computational}. Finding the ground state for a quantum system with a local Hamiltonian seems to be even harder; it is QMA-hard \cite{kitaev2002classical}, as hard as any problem whose solution can be checked efficiently by a quantum computer, and we expect that QMA is a larger class than NP. Surprisingly, computing the ground-state energy seems to be a hard problem for a quantum computer even for the case of a geometrically local translationally-invariant quantum system in one dimension \cite{gottesman2009quantum}. That is, it follows from rather weak complexity-theoretic assumptions that there are hard instances of the one-dimensional version of the ground-state-energy problem, even though it is often easy in cases of physical interest.

A general procedure for preparing ground states is adiabatic evolution. We can prepare a state having sizable overlap with the ground state of $H$ by starting with the easily prepared ground state of a simpler Hamiltonian $H(0)$, then slowly deforming the Hamiltonian along a path $H(s)$ connecting $H(0)$ to $H(1)=H$. This procedure succeeds in polynomial time provided the energy gap $\Delta(s)$ between the ground and first excited states of $H(s)$ is no smaller than inverse polynomial in $n$ for all $s\in[0,1]$ along the path. For problem instances that are quantumly hard, then, the gap becomes superpolynomially small somewhere along the path \cite{farhi2000quantum}.

Though the general problem is quantumly hard, we may surmise that there are many local quantum systems for which computing the ground-state energy is quantumly easy yet classically hard. 
%
For example, the electronic structure of a molecule with atomic nuclei at fixed positions can be accurately described by a local Hamiltonian,
and chemists assert (without proof) that it is possible to evolve adiabatically from the {\it Hartree-Fock} Hamiltonian (which they can solve classically) to the {\it full configuration interaction} (FCI) Hamiltonian (which they want to solve, but don't know how to solve classically in general), while the gap $\Delta$ exceeds a nonzero constant everywhere along the adiabatic path \cite{aspuru2005simulated}. If that is true, someday fully scalable fault-tolerant quantum computers will be powerful tools for advancing molecular chemistry.

\section{Quantum error correction}
\label{sec:qec}

Classical digital computers exist, and have had a transformative impact on our lives. Large-scale quantum computers do not yet exist. Why not?

Building reliable quantum hardware is challenging because of the difficulty of controlling quantum systems accurately. Small errors in quantum gates accumulate in a large circuit, eventually leading to large errors that foil the computation. Furthermore, qubits in a quantum computer inevitably interact with their surroundings; decoherence arising from unwanted correlations with the environment is harmless in a classical computer (and can even be helpful, by introducing friction which impedes accidental bit flips), but decoherence in a quantum computer can irreparably damage the delicate superposition states processed by the machine.

Quantum information can be better protected against noise by using a quantum error-correcting code, in which ``logical'' information is encoded redundantly in a block of many physical qubits \cite{shor1995scheme,steane1996error}. Quantum error correction is in some ways much like classical error correction, but more difficult, because while a classical code need only protect against bit flips, a quantum code must protect against both bits flips and phase errors. 

\subsection{Conditions for quantum error correction}\label{subsec:conditions}

Suppose for example, that we want to encode a single logical qubit, with orthonormal basis states denoted $|\bar 0\rangle$ and $|\bar1\rangle$, which is protected against all the errors spanned by a set $\{E_a\}$. For the distinguishability of the basis states to be maintained even when errors occur, we require
\begin{equation}\label{eq:bit-perp}
E_a|\bar 0\rangle \perp E_b|\bar1\rangle,
\end{equation}
where $E_a,E_b$ are any two elements of the error basis. This condition by itself would suffice for reliable storage of a classical bit.

But for storage of a qubit we also require protection against phase errors, which occur when information about whether the state is $|\bar 0\rangle$ or $|\bar1\rangle$ leaks to the environment; equivalently, distinguishability should be maintained for the dual basis states $\left(|\bar 0\rangle \pm |\bar 1\rangle\right)/\sqrt{2}$:
\begin{equation}\label{eq:phase-perp}
E_a\left(|\bar 0\rangle +|\bar1\rangle\right)\perp E_b\left(|\bar 0\rangle -|\bar 1\rangle\right),
\end{equation}
where $E_a,E_b$ are any two errors. In fact, the two distinguishability conditions eq.(\ref{eq:bit-perp}) and (\ref{eq:phase-perp}) suffice to ensure the existence of a recovery map that corrects any error spanned by $\{E_a\}$ acting on any linear combination of $|\bar 0\rangle$ and $|\bar 1\rangle$ \cite{knill1997theory}.

Together, eq.(\ref{eq:bit-perp}) and (\ref{eq:phase-perp}) imply
\begin{equation}\label{eq:bit-same}
\langle \bar 0| E_a^\dagger E_b |\bar 0\rangle = \langle \bar 1| E_a^\dagger E_b |\bar 1\rangle;
\end{equation}
no measurement of any operator in the set $\{E_a^\dagger E_b\}$ can distinguish the two basis states of the logical qubit. Typically, because we expect noise acting collectively on many qubits at once to be highly suppressed, we are satisfied to correct {\em low-weight} errors, those that act nontrivially on a sufficiently small fraction of all the qubits in the code block. Then eq.(\ref{eq:bit-same}) says that all the states of the logical qubit look the same when we examine a small subsystem of the code block. To be well protected, the logical states should be highly entangled, so that no logical information is accessible locally.

\begin{figure}
\begin{center}
\includegraphics[width=0.7\textwidth]{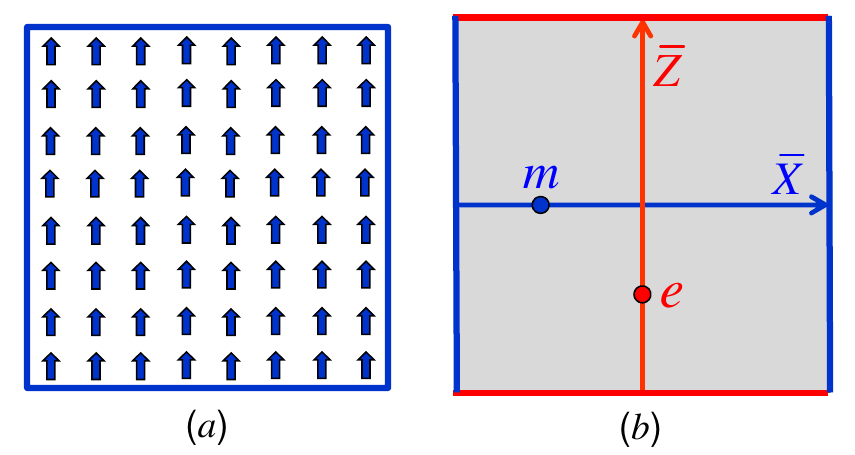}
\end{center}
\caption{\label{fig:memory} ($a$) A prototypical classical memory is a ferromagnet. ($b$) A prototypical quantum memory is a topologically ordered medium.}
\end{figure}

\subsection{Protected quantum memory and topological order}\label{subsec:protected}
\begin{figure}
\begin{center}
\includegraphics[width=\textwidth]{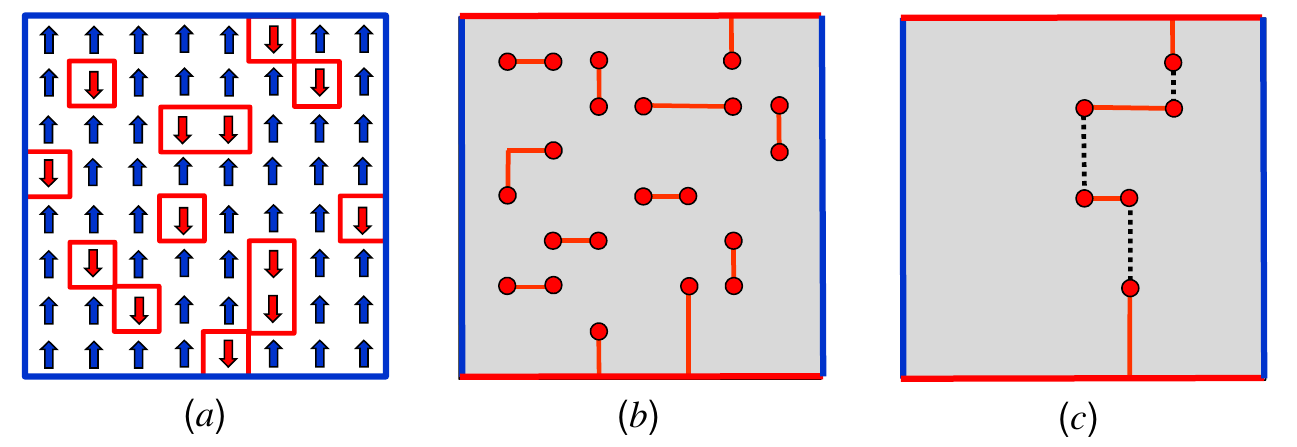}
\end{center}
\caption{\label{fig:defects} ($a$) In a two-dimensional ferromagnet, domain walls surround droplets of flipped bits. ($b$) In a two-dimensional topological quantum memory, pointlike anyons appear at the ends of chains of flipped qubits. ($c$) A logical error occurs if the actual errors (red) combine with our error diagnosis (dotted black) to produce a path that traverses the medium.  }
\end{figure}

It is useful to formulate the distinction between classical and quantum error correction in more physical terms (see Fig.~\ref{fig:memory} and \ref{fig:defects}). The prototype for a protected classical memory is a ferromagnet, where a single bit is encoded according to whether most of the spins are up or down. The encoded bit can be read out by performing local measurements on all spins, and then executing a majority vote to protect against errors that flip a minority of the spins. Errors in the memory create domain walls where neighboring spins misalign, and a logical error occurs when a domain wall sweeps across the sample, inducing a global operation acting on many spins. 
The memory is robust at a sufficiently small nonzero temperature because large droplets of flipped spins have a large energy cost, and are therefore unlikely to occur due to thermal fluctuations. This memory is a particularly simple physically motivated example of a classical error-correcting code; there are more sophisticated examples. 

The prototype for a protected quantum memory is a medium in two dimensions with $\mathbb{Z}_2$ topological order \cite{kitaev2003fault}. 
In contrast to the ferromagnet, errors in the medium create pointlike excitations (``anyons'') rather than domains walls. There are two types of anyons, which we denote as $e$ (for ``electric'') and $m$ (for ``magnetic''). We speak of ``$\mathbb{Z}_2$ topological order'' because when an $e$ anyon travels around an $m$ anyon (or an $m$ anyon travels around an $e$ anyon), the many-body wave function acquires a ``topological phase'' of -1. This phase is topological in the sense that it does not matter what path the $e$ anyon follows; all that matters is that it winds around the $m$ anyon an odd number of times. $\mathbb{Z}_2$ just means that the topological phase takes two possible values, +1 and -1. 

There are two possible types of one-dimensional edge for this two-dimensional medium, shown in Fig.~\ref{fig:memory}b. An $e$ anyon can appear or disappear at the red edge on the top and the bottom of the sample, while an $m$ anyon can appear or disappear at the blue edge on the left and on the right. The protected code space is the space of quantum states in which no anyons are present. There are nontrivial physical processes which preserve this code space. Namely, an $e$ anyon can appear at the bottom, propagate across the sample and disappear at the top. This process applies a unitary operator to system, which we call $\bar Z$. Or an $m$ anyon can appear at the left, propagate across, and disappear at the right. This process applies a different unitary operator to the system, which we call $\bar X$. Because of the topological phase -1 that arises when $e$ winds around $m$, these operators do not commute; rather
\begin{equation}
\bar X^{-1}\bar Z^{-1}\bar X \bar Z= - I,
\end{equation}
Thus two anticommuting operators both preserve the code space, which means the code space cannot be one dimensional. In fact it is two dimensional, and we may interpret $\bar Z$ and $\bar X$ as the Pauli operators acting on the protected qubit:
\begin{equation}
\bar  Z =\left(
 \begin{array}{cc}
 1 & 0\\ 0 & -1
 \end{array}
\right),
 \quad
 \bar X = \left(\begin{array}{cc}
 0 & 1\\ 1 & 0
 \end{array}
\right).
 \end{equation}
I have written a bar above $\bar Z$ and $\bar X$ to distinguish these ``logical'' Pauli operators, which act on the encoded qubit, from the physical Pauli operators we will discuss in Sec.~\ref{subsec:surface} below.
 
 There is another process we might consider. A pair of $e$ anyons (or $m$ anyons) are created in the bulk of the sample, away from any boundary. These anyons wander around for a while, without every approaching the boundary, until finally they find one another, annihilate, and disappear. This process also preserves the code space, but in contrast to the $\bar Z$ and $\bar X$ operators, it acts trivially on the protected qubit; that is, it commutes with both $\bar Z$ and $\bar X$. For example, the path followed by the $m$ anyon from the left to right edge can be deformed so it stays away from the pair of $e$ anyons wandering in the bulk. This does not change how $\bar X$ acts on the code space, but makes clear that the diffusing pair of $e$ anyons can't have any effect on $\bar X$.


The system is protected by a nonzero energy gap, the energy cost of creating a pair of anyons. Hence quantum information can be stored for a long time if the temperature is small compared to the gap, but unlike the case of the two-dimensional ferromagnet, the storage time does not improve as the system size increases. In the ferromagnet, the energy cost of a bubble of flipped spins increases as the bubble grows; in contrast, once a pair of anyons is thermally excited, no further energy barrier prevents the anyons from wandering to opposite sides of the sample, producing a logical $\bar Z$ or $\bar X$ error. However, if we \emph{monitor} the particles as they diffuse through the sample, then a logical error occurs only if particles propagate across the sample without being noticed, an event which {\em does} become increasingly unlikely as the system size grows \cite{dennis2002topological}. The scheme for performing robust quantum computation described in the Sec.~\ref{subsec:surface} builds on this observation.

\subsection{Surface code accuracy threshold}\label{subsec:surface}

To create a stable quantum memory, we need not synthesize a topologically ordered material; instead we can \emph{simulate} the material using whatever quantum computing hardware we prefer.
Kitaev constructed a simple two-dimensional lattice model (the \emph{surface code}), with a qubit at each lattice site, that exhibits $\mathbb{Z}_2$ topological order just as described in Sec.~\ref{subsec:protected} \cite{kitaev2003fault,bravyi1998quantum,freedman2001projective}. 
Though it was first proposed nearly 25 years ago, the surface code still offers a particularly promising route toward scalable fault-tolerant quantum computation. It has two major advantages. First, the quantum processing needed to diagnose and correct errors is remarkably simple. Second, and not unrelatedly, it can tolerate a relatively high gate error rate. 

Errors afflicting a quantum memory can be expanded in terms of multi-qubit Pauli operators, and each such Pauli operator can be expressed as a product of an $X$-type error, where either $X$ or the identity acts on each qubit, and a $Z$-type error, where either $Z$ or the identity acts on each qubit. (A $Y=-iZX$ error is just the case where both $X$ and $Z$ act on the same qubit.) 
Therefore, our quantum memory will be well protected if we can correct both $X$-type and $Z$-type errors with high success probability. In the case of the surface code, there are two separate procedures for correcting $X$ errors and correcting $Z$ errors, and both work in essentially the same way, so it will suffice to discuss only how the $Z$ errors are corrected. 


In (one version of) the surface code, the physical qubits reside on edges of a square lattice, and $e$ anyons may reside on the sites of the lattice. Suppose an unknown quantum state $\alpha |\bar 0\rangle + \beta |\bar 1\rangle$ has been stored in the code space, where $|\bar 0\rangle$ and $|\bar 1\rangle$ are the encoded $\bar Z$ eigenstates. After this state is encoded, $Z$ errors occur on some of the qubits, knocking the state out of the code space by creating $e$ anyons. A snapshot of a typical error configuration is shown in Fig.~\ref{fig:defects}b; edges on which the qubits have $Z$ errors (colored red), define a set of connected ``error chains,'' and pairs of anyons appear at the endpoints of each error chain. The positions of the anyons (and hence the endpoints of the error chains) can be identified by a simple quantum computation. After finding their positions we can remove these anyons two at a time; we select a pair of anyons, and apply $Z$ to all the qubits along a ``recovery chain'' that connects the pair, in effect bringing the pair of anyons together to annihilate.  Alternatively, we can remove a single anyon by choosing a recovery chain that connects that anyon to the top or bottom edge. Our goal is to remove all of the anyons, returning the state to the code space, and (we hope) restoring the initial encoded state. 

The anyon positions are said to constitute an error ``syndrome'' because they help us to diagnose the damage sustained by the physical qubits in the code block. Even though this syndrome locates the boundary points of the error chains, we don't know the configuration of the error chains themselves, so our recovery chains won't necessarily coincide with the error chains, or even connect together the same pairs of anyons. But they don't have to. If each connected path resulting from combining the error chain with the recovery chain forms a closed loop in the bulk, or an open path with both its endpoints lying on the same edge (either top or bottom), then error recovery is successful. This works because of the properties of the topologically ordered medium noted earlier: creation of a pair of anyons followed by pair annihilation, or creation of a single anyon at the bottom (top) edge followed by annihilation at the bottom (top) edge are processes that act trivially on the code space. On the other hand, if the error chain combined with the recovery chain produces a path connecting the bottom and top edges as in Fig~\ref{fig:defects}c, then (if there are an odd number of such paths) a logical $\bar Z$ error occurs and our recovery procedure fails. 

To keep things simple, consider a stochastic independent noise model, in which each qubit in the code block experiences a $Z$ error with probability $\epsilon$. Suppose we choose our recovery chains to have the minimal possible weight; that is, we return to the code space by applying $Z$ to as few qubits as possible. Given the known positions of the anyons, this minimal chain can be computed efficiently with a classical computer. For this recovery procedure, we can find an upper bound on the probability of a logical error by the following argument \cite{dennis2002topological}. 

We denote by $d$ the minimal weight of a connected path from the bottom edge to the top edge; that is, $d$ (the \emph{distance} of the code) is the minimal weight of a $\bar Z$ logical operator. If our attempt to recover resulted in a logical $\bar Z$ error, there must be a path connecting the bottom and top edges of the code block such that each edge of the lattice on this path is in either an error chain or a recovery chain. Let's say this connected path has length $\ell\ge d$, and denote the path by $C_\ell$. The number of errors on $C_\ell$ must be at least $\ell/2$ if $\ell$ is even, or $(\ell+1)/2$ if $\ell$ is odd; otherwise we could have found a lower weight recovery chain by applying $Z$ on the error chains contained in $C_\ell$, rather than to the qubits on $C_\ell$ which are complementary to the error chains on $C_\ell$. The number of ways that the edges with $Z$ errors could be distributed along $C_\ell$ is no more than $2^{\ell}$ (each qubit on $C_\ell$ either has an error or does not). Since, for each physical qubit, $Z$ errors occur with probability $\epsilon$, the probability that $C_\ell$ is contained in the union of error chains and recovery chains obeys
\begin{equation}\label{eq:C-ell-bound}
P(C_\ell) \le 2^\ell \epsilon^{\ell/2}.
\end{equation}
Let $N_\ell$ denote the number of paths connecting the bottom and top edges with length $\ell$. For a logical error to occur, the combination of error chains and recovery chains must produce at least one path connecting the bottom and top edges. Using the upper bound eq.(\ref{eq:C-ell-bound}) on the probability  of each such path, and applying the union bound, we conclude that the probability of a logical $\bar Z$ error satisfies
\begin{equation}
P_{\rm logical} \le \sum_{\ell = d}^n N_\ell 2^\ell \epsilon^{\ell/2}.
\end{equation}
The lower limit on the sum is $\ell=d$, the length of the shortest path connecting the bottom and top edges. The upper limit is $n$, the total number of qubits in the code block, which is therefore the maximum length of any path. 

We can also find a simple upper bound on $N_\ell$. Let's say our square lattice is $d\times d$. A path from the bottom to the top edge can begin at any one of $d$ positions along the bottom edge, and in each of the $\ell$ steps along the path, there are three possible moves: straight ahead, left turn, or right turn. Therefore (even if we don't insist that the path reach the top edge), we have
\begin{equation}
N_\ell\le d ~3^{\ell}\implies P_{\rm logical} \le d \sum_{\ell = d}^n  \left( 36\epsilon\right)^{\ell/2}.
\end{equation}
Now suppose that $\epsilon < 1/36$, so that the terms in the sum over $\ell$ decrease as $\ell$ increases. For a square lattice, the number of edges (qubits) in the code block is $n=O(d^2)$, so the number of terms in the sum is also $O(d^2)$, and we conclude that
\begin{equation}
 P_{\rm logical} \le O(d^3)   \left(\epsilon/\epsilon_0\right)^{d/2}\quad \mathrm{for}\quad \epsilon < \epsilon_0= 1/36\approx .028~.
\end{equation}
Thus, this argument establishes that the surface code is a quantum memory with an \emph{accuracy threshold} --- for any constant $\epsilon < \epsilon_0$, the probability of a logical error decays exponentially as the code distance $d$ increases (apart from a possible polynomial prefactor). If the physical error rate is below the threshold value $\epsilon_0$, we can make the logical error rate arbitrarily small by choosing a sufficiently large code block. Unsurprisingly, in view of the crudeness of this argument, the actual value of the error threshold $\epsilon_0$ is larger than we estimated. Monte Carlo simulations find $\epsilon_0\approx .103$ \cite{wang2003confinement}.

\subsection{Scalable quantum computing}

To draw quantitative conclusions about the overhead cost of fault-tolerant quantum computing, refinements of this argument are needed. First of all, we implicitly assumed that the error syndrome measurements are perfect. In fact measurement errors occur, which means we need to repeat the measurement $O(d)$ times to acquire sufficiently trustworthy information about where the anyons are located. Secondly, we did not take into account the structure of the quantum circuit used to make these measurements. To determine whether an anyon is present at a particular site, four entangling two-qubit gates are needed, any one of which could be faulty, and a single fault can cause both an error in the measurement outcome and errors in the data qubits. A more complete analysis shows that the threshold error rate for the two-qubit gates is close to 1\% \cite{raussendorf2007fault,raussendorf2007topological}. Numerical simulations find that for each round of syndrome measurement, the probability of a logical error rate scales roughly like \cite{Fowler_2013} 
\begin{equation}\label{eq:p-logical}
P_{\rm logical} \approx 0.1  \left(100 p\right)^{(d+1)/2},
\end{equation}
where now $p$ denotes the two-qubit gate error rate, and we have assumed that $d$ is odd. 

So far we have considered only the probability of a storage error for one protected qubit, but in a scalable fault-tolerant quantum computer we will need many protected qubits, and we will need to perform highly reliable universal quantum gates that act on these qubits. One can envision an architecture in which the logical qubits are arranged like square tiles on a surface, with buffer qubits filling gaps between the tiles \cite{fowler2018low,litinski2019game,litinski2019magic}. I won't go into the details of how the logical gates are executed, but it is helpful to realize that much of the logical processing can be executed by performing entangling measurements on pairs of logical blocks. For example, we can measure $\bar X_1\otimes \bar X_2$, where blocks 1 and 2 reside on adjacent tiles, by fusing the blocks together along their red edges and then cutting the blocks apart again, a process called ``lattice surgery'' \cite{horsman2012surface}. The fusing and cutting are achieved by measurements that activate the buffer qubits in between the edges of the two blocks, followed by measurements that decouple the buffer qubits. 

The good news is that the error rates for logical gates are not much worse than the storage error rates we have already discussed, except we should keep in mind that we need to repeat the syndrome measurement $O(d)$ times in each logical gate cycle. The bad news is that eq.(\ref{eq:p-logical}) indicates that we'll need a rather large code distance if we want to make the logical error rate very small. Suppose, for example, that we would like to run Shor's algorithm to factor a 2048-bit number, which would break the RSA cryptosystem, and suppose that the physical two-qubit gate error rate is $10^{-3}$, better than in current multi-qubit devices. The analysis in \cite{gidney2021factor} calls for a logical error probability $\approx 10^{-15}$ per round of syndrome measurement, and hence a code distance of $d=27$. The number of physical qubits per code block, including ancilla qubits needed for syndrome measurement and lattice surgery, is $2(d+1)^2= 1568$, and the total number of logical qubits used in this version of the factoring algorithm is about 14,000, pushing the physical qubit count above 20 million \cite{gidney2021factor}. That's a lot!

There are many challenges to making large-scale fault-tolerant quantum computing practical, including serious systems engineering issues. There are also issues of principle to consider --- what is required for a fault-tolerant scheme to be scalable, and what conditions must be satisfied by the noise model? One essential requirement is some form of cooling, to extract the entropy introduced by noise \cite{aharonov1996limitations}. In the protocol described above, entropy is extracted by measuring and resetting ancilla qubits in each round of syndrome measurement. Parallel operations are also necessary, so noise can be controlled in different parts of the computer simultaneously.

The analysis leading to eq.(\ref{eq:p-logical}) is based on a simple noise model in which gate errors are stochastic (rather than coherent) and there are no correlations among errors in different gates. The fault-tolerant methods should work for more realistic noise models, as long as the errors are sufficiently weak and not too strongly correlated. By benchmarking logical error rates using relatively small quantum codes during the NISQ era, we will gain valuable insight into how effectively quantum error correction protects computations performed on actual quantum hardware.

\section{Outlook}
\label{sec:outlook}

I cherish the memory of some very enjoyable conversations with Dick Feynman about physics and about other things, too. But quantum computing was one subject we never discussed. I knew Feynman was interested in quantum computation, but I was not very interested back then, so I never asked him about it. Naturally I regret that now. Six years after his death I became very interested, but by then it was too late. 

The key issues we might have discussed in the early 1980s still loom over the subject today.
Can we build powerful large-scale quantum computers? How will we do it? When will it happen? And what will we do with those awesome machines? I'm confident that the answer to the first question is yes. But 40 years later the answers to the other questions are still far from clear. Feynman was right to conclude his 1981 talk with the observation, ``it’s a wonderful problem because it doesn’t look so easy.''

Those who aspire to build quantum computing systems face a daunting engineering challenge, but there's more to it than that. It would transform the prospects for practical quantum computing applications if we could improve physical gate error rates (currently around 1\% for entangling two-qubit gates) by several orders of magnitude. The progress achieved so far has been driven by advances in qubit design, control technology, fabrication methods, and materials, and further incremental improvements can surely be expected. But quantum hardware is still at an early stage, and truly disruptive progress might flow from fresh ideas about how to encode and manipulate quantum information robustly. The quantum community should continue to think broadly and imaginatively about new approaches to building and operating quantum hardware. 

Feynman was on the right track when he suggested using quantum computers to solve problems in quantum physics and chemistry. That is still the most important application we can clearly foresee, and there is plenty of opportunity to flesh out our ideas about how quantum computers can best be used to advance science. Applications of broader interest are also possible. Quantum computers can speed up exhaustive search for solutions to optimization problems, but because the speedup is only quadratic in that case, this might not be useful until far in the future. More dramatic quantum speedups for optimization and related problems cannot yet be ruled out, and should continue to be a goal for research on quantum algorithms. 

Though fully scalable fault-tolerant quantum computers may still be a ways off, the advent of the NISQ era already heralds unprecedented opportunities for exploring the properties of highly entangled many-body quantum systems. With NISQ technologies, we will also assess the performance of heuristic hybrid quantum/classical algorithms, which may steer us toward practical applications, and we'll advance our toolkit for mitigating noise and correcting errors in quantum platforms. Today's quantum computers can help us to build tomorrow's more powerful quantum computers. 

This chapter has focused on the prospects for building and using quantum computers. But if Feynman were here today, I would be just as eager to tell him about the myriad of ways that quantum information concepts have opened new vistas across many domains of physics. To mention two prominent examples, we have understood that different quantum phases of matter can be distinguished according to the structure of their long-range quantum entanglement \cite{zeng2019quantum}, and that the spacetime geometry in a model of quantum gravity can admit an alternative description, in which the geometry is encoded in the quantum entanglement of a quantum system that does not involve gravitation at all \cite{ryu2006holographic}. Powerful insights like these signal that quantum information science has become an essential force in humanity's struggle to grasp Nature's hidden secrets. From now on, quantum computer science and quantum physical science will advance together, hand in hand.

\section{Memories of Feynman at Caltech}
\label{sec:memories}

\subsection{Getting acquainted}

Richard Feynman and I overlapped at Caltech for $4\frac{1}{2}$ years, from the start of my faculty appointment in August 1983 until Feynman's death in February 1988.  Our relationship as colleagues got off to a great start. 

One day soon after I arrived, I hear someone drumming on the wall while walking down the hallway, know it must be Feynman, and step out of my office to greet him. 

\begin{description}

\item Our theory group admin Helen Tuck introduces us: ``Dr. Feynman, this is Dr. Preskill, our new faculty member!'' 

\item Feynman replies:``What group?'' 

\item Does Feynman really not know who I am? “Um … particle theory.” 

\item Feynman: “People who say they do particle theory do many different things. What do you do?” 

\item I ramble incoherently for a minute about the connection between particle physics and cosmology, then unwisely conclude:``And lately, I have been working, without much success, on models in which quarks and leptons are composite.'' 

\item Long pause, then: ``Well, your lack of success has been shared by many others.'' Feynman turns and disappears into his office. 

\end{description}

So right away I knew we would be friends.

\subsection{Caltech seminar}
Actually, Feynman and I had already met a few times before that, when I had come to Caltech to give seminars. Speaking at the Caltech particle theory seminar in the days of Feynman and Gell-Mann was a memorable experience. Here is how Steve Weinberg described it \cite{weinberg1998}:

\begin{quote}
Years ago, when I was an assistant professor of physics at Berkeley [1960-66], I used to be invited down to Caltech about once a year to give a talk. It was usually the low point of my year. In the audience at Caltech were two leaders of modern physics, Murray Gell-Mann and Richard Feynman, who interrupted with frequent questions, ruthlessly probing to see if I really knew what I was talking about and had anything new to say. Of the two, Feynman was the more frightening. Gell-Mann was mostly interested in finding out whether there was anything in my talk that he should know about, so he was no problem if I did have anything worthwhile to say. Feynman was having fun.
\end{quote}

\noindent 15 years later, when I gave my first seminar at Caltech, the experience was not quite so terrifying. By then, I discovered, one could play Feynman and Gell-Mann against one another. When Dick attacked, Murray defended me. And when Murray raised an objection, Dick would be on my side. It was an eventful seminar, but not ``the low point of my year.''

\subsection{The World of Science}

My first ``encounter'' with Feynman had actually occurred years earlier, when at age 9 I acquired a marvelous book called \emph{The World of Science} by Jane Werner Watson \cite{watson1958world}. There was a chapter about theoretical physics, which began with the story of a boy whose little red wagon has a ball in the back. The boy notices that when he pulls the wagon forward the ball rolls backward, and when he stops pulling the ball rolls forward. He asks his father why, and his father replies: ``That’s called inertia, but nobody knows why.''

Some 20 years later I watched the terrific interview of Feynman by Christopher Sykes, called \emph{The Pleasure of Finding Things Out}, where Feynman tells the same story \cite{sykes1981pleasure}. Whoa!, I thought --- did Feynman steal the story from the Golden Book I read as a child? Looking at the book for the first time in years, I realized what had happened. Children's author Jane Werner Watson was married to Earnest Watson, the Caltech Dean of the Faculty, and she based the book on interviews with faculty members. 

What I found particularly inspiring was the discussion in that book of a discovery made just a year before the book was published --- that the laws of physics governing elementary particles know the difference between left and right! That amazing fact kindled a passion for physics that eventually carried me to Caltech, where I joined Dick Feynman on the faculty 21 years later. 

\subsection{Talking physics}

As Caltech colleagues, Feynman and I found that we had a common interest in nonperturbative aspects of quantum chromodynamics, in particular why quarks are confined inside hadrons, and we often talked about that. Sometimes I would impress Feynman with an idea or calculation I had learned from the literature; I would tell him the source, but he would be interested in the ideas, not the reference. Once I overheard Feynman tell Helen, as he returned to his office after our discussion: ``He’s like an encyclopedia. No, he’s better than the encyclopedia!'' That made my day. But sometimes I wondered whether Feynman knew my name, as he sometimes seemed to confuse ``Preskill'' and ``[Michael] Peskin'' --- with whom, I presumed, he had discussed similar things.  

Feynman and Gell-Mann had once been close, but there was evident tension between them by the 1980s. After I had gotten to know them better, I asked each one what had gone wrong. Both gave the same answer --- they had gotten along well until around 1969, when Feynman was working on the parton model. Years later, Murray spoke derisively about Feynman’s ``put-ons,'' still resentful that Feynman had refused to call them ``quarks.'' Feynman for his part, recalled that Gell-Mann had scathingly ridiculed the idea that quarks would behave as nearly free particles inside hadrons. What started out as a scientific disagreement had become increasingly personal and hostile, and their relationship never fully recovered. 

\subsection{Feynman's belt trick}
David Goodstein once asked Feynman to explain to him something rather esoteric in the theory of elementary particles, what we call the connection between spin and statistics \cite{goodstein1989richard}. Feynman promised to prepare a freshman lecture on it. But sometime later he returned and sheepishly admitted: ``I couldn't do it. I couldn't reduce it to the freshman level. That means we don't really understand it.''

But later he changed his mind. 
And Feynman, with great relish, showed me his way of explaining spin and statistics. I’ve been using it ever since, whenever I teach our sophomore class at Caltech, Physics 12. 
It goes like this. 

First he boiled down the problem to its essence. The connection between spin and statistics means this: that when two elementary particles, two electrons say, change places, that’s really the same thing as having one of the particles spin around by 360 degrees. And he’d illustrate what it means to spin around by doing a pirouette. Then he would take off his belt. And he’d say: ``Look. The two ends of my belt are two electrons. What happens when I exchange them? At first you think nothing has happened to the belt. but wait, it has a twist in it! Now let’s twist one end of it 360 degrees. See, the belt is the same as when it started. So exchanging the two electrons, and then twisting one of them, is that same as doing nothing! That’s the connection between spin and statistics.''

It's a visceral demonstration. And for the past 35 years, I’ve been thinking about how to interpret that belt. 

In 1987, just months before he died, Feynman took a group of students to dinner at a local restaurant, where somehow the subject of spin and statistics came up. And so, with as much relish as ever, Feynman stood up from the table, and started taking off his belt. Just at that moment, the waitress came up to the table, took in what was going on, shook her finger and scolded: ``That’s as far as it goes!''

\subsection{The final blackboard}
When Feynman died in 1988, the blackboard in his office was photographed, and that photo was widely disseminated. Two prominent passages appear: “What I cannot create, I do not understand” and “Know how to solve every problem that has been solved.” Why those words? I may be able to shed some light on that. 

In late 1986, after the Rogers Commission had finished their investigation of the Challenger disaster, Feynman was eager to dive back into physics, and particularly excited about further investigations of QCD. Feynman was interested in lattice QCD, which he recognized as a beautiful application of path integral methods (which he had invented decades earlier \cite{feynman2010quantum}), but he felt that computational power was then inadequate for getting accurate results, and would remain inadequate for some time to come. Meanwhile, he hoped to make progress with analytic methods, or a combination of numerical and analytic methods. 

In particular, he hoped that tools for solving integrable models might be helpful for treating the soft part of QCD, the physics beyond the reach of renormalization-group improved perturbation theory. He wanted some students to study integrable models with him, to help him learn the subject. Well, he wanted students, and I had students, so we made an arrangement. Feynman and the students met once a week in his office, and those meetings would sometimes last all afternoon; a few times Feynman invited the students to dinner afterward. 

Feynman told the students ``We gotta know how to solve every problem that has been solved,'' and he urged them to solve the problems on their own because ``What I cannot create I do not understand.'' To get things started he described the six-vertex model, and told everyone to solve it without looking up any references \cite{baxter2016exactly}. That went on for weeks, without notable progress, until Feynman triumphantly unveiled his own solution. The next challenge was the eight-vertex model, but the students never solved that one, and neither did Feynman!

One of the students was Sandip Trivedi, who recalls that Feynman was becoming ill, but was ``incredibly enthusiastic and extremely patient'' with the students. During his final illness, he told Helen to share his notes with the students, and they were all amazed and inspired to see how meticulous and detailed the notes were, containing many intricate calculations.

We lost Richard Feynman on February 15, 1988. It was a very sad day at Caltech. Feynman was loved, admired, and held in awe by a large swath of the campus community; he was part of the soul of the place, and he’s still sorely missed by those who knew him. Since then, many students have come and gone who never knew Feynman personally, but they too have been deeply influenced by his contributions, writings, ideas, and unique personality. The spirit of Richard Feynman lives on among deeply curious people everywhere. 

\subsection{A poem for Feynman}
I hope that readers of \emph{Surely You're Joking, Mr. Feynman!} \cite{feynman2018surely,feynman2001you} and others familiar with Feynman's life and career will appreciate this poem I wrote in honor of his 100th birthday, on May 11, 2018.

\begin{quote}
The Feynman legend, pundits say \\
Began in Queens – Far Rockaway. \\
It’s there a boy would stop and think \\
To fix a radio on the blink. \\
\\
He grew up as a curious guy \\
Who showed his sister the night sky. \\
He wondered why, and wondered \emph{why} \\
He wondered why he wondered why. \\
\\
New Jersey followed MIT. \\
The cream and lemon in his tea \\
Taught Mr. Feynman when to joke \\
And how to act like normal folk. \\
\\
Cracking safes, though loads of fun, \\
Could not conceal from everyone, \\
The mind behind that grinning brow: \\
A new Dirac, but human now. \\
\\
In New York state he spun a plate \\
Which led, in nineteen forty-eight \\
To diagrams that let us see \\
The processes of QED. \\
\\
He left the east and made a trek \\
Until he landed at Caltech. \\
His genius brought us great acclaim. \\
This place would never be the same. \\
\\
Dick’s teaching skills were next to none \\
When reinventing Physics 1. \\
His wisdom’s there for all to see \\
In red books numbered 1, 2, 3. \\
\\
Always up and never glum \\
He loved to paint and play the drum. \\
His mind engaged with everything \\
For all the world is int’resting. \\
\\
Dick proved that charm befits a nerd. \\
For papers read, and stories heard \\
We’ll always be in Feynman’s debt. \\
A giant we cannot forget. \\

\end{quote}

\bibliographystyle{unsrt}
\bibliography{feynman}

\begin{thebibliography}{10}

\bibitem{feynman21simulating}
Richard~P Feynman.
\newblock Simulating physics with computers.
\newblock {\em International Journal of Theoretical Physics}, 21:467--488,
  1981.

\bibitem{manin1980computable}
Yuri Manin.
\newblock Computable and {Uncomputable}.
\newblock {\em Sovetskoye Radio, Moscow}, 128, 1980.

\bibitem{benioff1980computer}
Paul Benioff.
\newblock The computer as a physical system: A microscopic quantum mechanical
  {Hamiltonian} model of computers as represented by turing machines.
\newblock {\em Journal of Statistical Physics}, 22(5):563--591, 1980.

\bibitem{deutsch1985quantum}
David Deutsch.
\newblock Quantum theory, the {Church--Turing} principle and the universal
  quantum computer.
\newblock {\em Proceedings of the Royal Society of London. A. Mathematical and
  Physical Sciences}, 400(1818):97--117, 1985.

\bibitem{bernstein1997quantum}
Ethan Bernstein and Umesh Vazirani.
\newblock Quantum complexity theory.
\newblock {\em SIAM Journal on Computing}, 26(5):1411--1473, 1997.

\bibitem{simon1997power}
Daniel~R Simon.
\newblock On the power of quantum computation.
\newblock {\em SIAM journal on Computing}, 26(5):1474--1483, 1997.

\bibitem{shor1999polynomial}
Peter~W Shor.
\newblock Polynomial-time algorithms for prime factorization and discrete
  logarithms on a quantum computer.
\newblock {\em SIAM review}, 41(2):303--332, 1999.

\bibitem{landauer1995quantum}
Rolf Landauer.
\newblock Is quantum mechanics useful?
\newblock {\em Philosophical Transactions of the Royal Society of London.
  Series A: Physical and Engineering Sciences}, 353(1703):367--376, 1995.

\bibitem{unruh1995maintaining}
William~G Unruh.
\newblock Maintaining coherence in quantum computers.
\newblock {\em Physical Review A}, 51(2):992, 1995.

\bibitem{haroche1996quantum}
Serge Haroche and Jean-Michel Raimond.
\newblock Quantum computing: dream or nightmare?
\newblock {\em Physics Today}, 49(8):51--54, 1996.

\bibitem{shor1995scheme}
Peter~W Shor.
\newblock Scheme for reducing decoherence in quantum computer memory.
\newblock {\em Physical Review A}, 52(4):R2493, 1995.

\bibitem{steane1996error}
Andrew~M Steane.
\newblock Error correcting codes in quantum theory.
\newblock {\em Physical Review Letters}, 77(5):793, 1996.

\bibitem{shor1996fault}
Peter~W Shor.
\newblock Fault-tolerant quantum computation.
\newblock In {\em Proceedings of 37th Conference on Foundations of Computer
  Science}, pages 56--65. IEEE, 1996.

\bibitem{aharonov1997fault}
Dorit Aharonov and Michael Ben-Or.
\newblock Fault-tolerant quantum computation with constant error.
\newblock In {\em Proceedings of the twenty-ninth annual ACM {Symposium on
  Theory of Computing}}, pages 176--188, 1997.

\bibitem{knill1998resilient}
Emanuel Knill, Raymond Laflamme, and Wojciech~H Zurek.
\newblock Resilient quantum computation.
\newblock {\em Science}, 279(5349):342--345, 1998.

\bibitem{kitaev1997quantum}
Aleksei~Yur'evich Kitaev.
\newblock Quantum computations: algorithms and error correction.
\newblock {\em Uspekhi Matematicheskikh Nauk}, 52(6):53--112, 1997.

\bibitem{preskill1998reliable}
John Preskill.
\newblock Reliable quantum computers.
\newblock {\em Proceedings of the Royal Society of London. Series A:
  Mathematical, Physical and Engineering Sciences}, 454(1969):385--410, 1998.

\bibitem{preskill1998fault}
John Preskill.
\newblock Fault-tolerant quantum computation.
\newblock In {\em Introduction to quantum computation and information}, pages
  213--269. World Scientific, 1998.

\bibitem{cirac1995quantum}
Juan~I Cirac and Peter Zoller.
\newblock Quantum computations with cold trapped ions.
\newblock {\em Physical Review Letters}, 74(20):4091, 1995.

\bibitem{monroe1995demonstration}
Chris Monroe, David~M Meekhof, Barry~E King, Wayne~M Itano, and David~J
  Wineland.
\newblock Demonstration of a fundamental quantum logic gate.
\newblock {\em Physical Review Letters}, 75(25):4714, 1995.

\bibitem{preskill1998quantum}
John Preskill.
\newblock Quantum computing: pro and con.
\newblock {\em Proceedings of the Royal Society of London. Series A:
  Mathematical, Physical and Engineering Sciences}, 454(1969):469--486, 1998.

\bibitem{feynman1960there}
Richard~P Feynman.
\newblock There's plenty of room at the bottom.
\newblock {\em California Institute of Technology, Engineering and Science
  magazine}, pages 22--36, 1960.

\bibitem{bernstein2017post}
Daniel~J Bernstein and Tanja Lange.
\newblock Post-quantum cryptography.
\newblock {\em Nature}, 549(7671):188--194, 2017.

\bibitem{wiesner1983conjugate}
Stephen Wiesner.
\newblock Conjugate coding.
\newblock {\em ACM Sigact News}, 15(1):78--88, 1983.

\bibitem{bennett2020quantum}
Charles~H Bennett and Gilles Brassard.
\newblock Quantum cryptography: Public key distribution and coin tossing.
\newblock {\em arXiv preprint arXiv:2003.06557}, 2020.

\bibitem{mosca2018cybersecurity}
Michele Mosca.
\newblock Cybersecurity in an era with quantum computers: will we be ready?
\newblock {\em IEEE Security \& Privacy}, 16(5):38--41, 2018.

\bibitem{mcardle2020quantum}
Sam McArdle, Suguru Endo, Alan Aspuru-Guzik, Simon~C Benjamin, and Xiao Yuan.
\newblock Quantum computational chemistry.
\newblock {\em Reviews of Modern Physics}, 92(1):015003, 2020.

\bibitem{bennett1997strengths}
Charles~H Bennett, Ethan Bernstein, Gilles Brassard, and Umesh Vazirani.
\newblock Strengths and weaknesses of quantum computing.
\newblock {\em SIAM Journal on Computing}, 26(5):1510--1523, 1997.

\bibitem{grover1997quantum}
Lov~K Grover.
\newblock Quantum mechanics helps in searching for a needle in a haystack.
\newblock {\em Physical Review Letters}, 79(2):325, 1997.

\bibitem{arute2019quantum}
Frank Arute, Kunal Arya, Ryan Babbush, Dave Bacon, Joseph~C Bardin, Rami
  Barends, Rupak Biswas, Sergio Boixo, Fernando~GSL Brandao, David~A Buell,
  et~al.
\newblock Quantum supremacy using a programmable superconducting processor.
\newblock {\em Nature}, 574(7779):505--510, 2019.

\bibitem{preskill2012quantum}
John Preskill.
\newblock Quantum computing and the entanglement frontier.
\newblock {\em arXiv preprint arXiv:1203.5813}, 2012.

\bibitem{harrow2017quantum}
Aram~W Harrow and Ashley Montanaro.
\newblock Quantum computational supremacy.
\newblock {\em Nature}, 549(7671):203--209, 2017.

\bibitem{huang2020classical}
Cupjin Huang, Fang Zhang, Michael Newman, Junjie Cai, Xun Gao, Zhengxiong Tian,
  Junyin Wu, Haihong Xu, Huanjun Yu, Bo~Yuan, et~al.
\newblock Classical simulation of quantum supremacy circuits.
\newblock {\em arXiv preprint arXiv:2005.06787}, 2020.

\bibitem{preskill2018quantum}
John Preskill.
\newblock Quantum computing in the {NISQ} era and beyond.
\newblock {\em Quantum}, 2:79, 2018.

\bibitem{farhi2014quantum}
Edward Farhi, Jeffrey Goldstone, and Sam Gutmann.
\newblock A quantum approximate optimization algorithm.
\newblock {\em arXiv preprint arXiv:1411.4028}, 2014.

\bibitem{peruzzo2014variational}
Alberto Peruzzo, Jarrod McClean, Peter Shadbolt, Man-Hong Yung, Xiao-Qi Zhou,
  Peter~J Love, Al{\'a}n Aspuru-Guzik, and Jeremy~L O’brien.
\newblock A variational eigenvalue solver on a photonic quantum processor.
\newblock {\em Nature Communications}, 5(1):1--7, 2014.

\bibitem{jaksch1998cold}
Dieter Jaksch, Christoph Bruder, Juan~Ignacio Cirac, Crispin~W Gardiner, and
  Peter Zoller.
\newblock Cold bosonic atoms in optical lattices.
\newblock {\em Physical Review Letters}, 81(15):3108, 1998.

\bibitem{greiner2002quantum}
Markus Greiner, Olaf Mandel, Tilman Esslinger, Theodor~W H{\"a}nsch, and
  Immanuel Bloch.
\newblock Quantum phase transition from a superfluid to a {Mott} insulator in a
  gas of ultracold atoms.
\newblock {\em Nature}, 415(6867):39--44, 2002.

\bibitem{bernien2017probing}
Hannes Bernien, Sylvain Schwartz, Alexander Keesling, Harry Levine, Ahmed
  Omran, Hannes Pichler, Soonwon Choi, Alexander~S Zibrov, Manuel Endres,
  Markus Greiner, et~al.
\newblock Probing many-body dynamics on a 51-atom quantum simulator.
\newblock {\em Nature}, 551(7682):579--584, 2017.

\bibitem{zhang2017observation}
Jiehang Zhang, Guido Pagano, Paul~W Hess, Antonis Kyprianidis, Patrick Becker,
  Harvey Kaplan, Alexey~V Gorshkov, Z-X Gong, and Christopher Monroe.
\newblock Observation of a many-body dynamical phase transition with a 53-qubit
  quantum simulator.
\newblock {\em Nature}, 551(7682):601--604, 2017.

\bibitem{chiu2019string}
Christie~S Chiu, Geoffrey Ji, Annabelle Bohrdt, Muqing Xu, Michael Knap, Eugene
  Demler, Fabian Grusdt, Markus Greiner, and Daniel Greif.
\newblock String patterns in the doped hubbard model.
\newblock {\em Science}, 365(6450):251--256, 2019.

\bibitem{mukherjee2019spectral}
Biswaroop Mukherjee, Parth~B Patel, Zhenjie Yan, Richard~J Fletcher, Julian
  Struck, and Martin~W Zwierlein.
\newblock Spectral response and contact of the unitary fermi gas.
\newblock {\em Physical Review Letters}, 122(20):203402, 2019.

\bibitem{semeghini2021probing}
Giulia Semeghini, Harry Levine, Alexander Keesling, Sepehr Ebadi, Tout~T Wang,
  Dolev Bluvstein, Ruben Verresen, Hannes Pichler, Marcin Kalinowski, Rhine
  Samajdar, et~al.
\newblock Probing topological spin liquids on a programmable quantum simulator.
\newblock {\em arXiv preprint arXiv:2104.04119}, 2021.

\bibitem{gottesman2010introduction}
Daniel Gottesman.
\newblock An introduction to quantum error correction and fault-tolerant
  quantum computation.
\newblock In {\em Quantum information science and its contributions to
  mathematics, Proceedings of Symposia in Applied Mathematics}, volume~68,
  pages 13--58, 2010.

\bibitem{campbell2017roads}
Earl~T Campbell, Barbara~M Terhal, and Christophe Vuillot.
\newblock Roads towards fault-tolerant universal quantum computation.
\newblock {\em Nature}, 549(7671):172--179, 2017.

\bibitem{kivlichan2020improved}
Ian~D Kivlichan, Craig Gidney, Dominic~W Berry, Nathan Wiebe, Jarrod McClean,
  Wei Sun, Zhang Jiang, Nicholas Rubin, Austin Fowler, Al{\'a}n Aspuru-Guzik,
  et~al.
\newblock Improved fault-tolerant quantum simulation of condensed-phase
  correlated electrons via trotterization.
\newblock {\em Quantum}, 4:296, 2020.

\bibitem{campbell2021early}
Earl~T. Campbell.
\newblock Early fault-tolerant simulations of the {Hubbard} model.
\newblock {\em arXiv preprint arXiv:2012.09238}, 2021.

\bibitem{holevo1973bounds}
Alexander~Semenovich Holevo.
\newblock Bounds for the quantity of information transmitted by a quantum
  communication channel.
\newblock {\em Problemy Peredachi Informatsii}, 9(3):3--11, 1973.

\bibitem{kitaev2002classical}
Alexei~Yu Kitaev, Alexander Shen, Mikhail~N Vyalyi, and Mikhail~N Vyalyi.
\newblock {\em {Classical and Quantum Computation}}.
\newblock Number~47. American Mathematical Soc., 2002.

\bibitem{jordan2012quantum}
Stephen~P Jordan, Keith~SM Lee, and John Preskill.
\newblock Quantum algorithms for quantum field theories.
\newblock {\em Science}, 336(6085):1130--1133, 2012.

\bibitem{preskill2018simulating}
John Preskill.
\newblock Simulating quantum field theory with a quantum computer.
\newblock {\em arXiv preprint arXiv:1811.10085}, 2018.

\bibitem{divincenzo2000physical}
David~P DiVincenzo.
\newblock The physical implementation of quantum computation.
\newblock {\em Fortschritte der Physik: Progress of Physics},
  48(9-11):771--783, 2000.

\bibitem{bruzewicz2019trapped}
Colin~D Bruzewicz, John Chiaverini, Robert McConnell, and Jeremy~M Sage.
\newblock Trapped-ion quantum computing: Progress and challenges.
\newblock {\em Applied Physics Reviews}, 6(2):021314, 2019.

\bibitem{kjaergaard2020superconducting}
Morten Kjaergaard, Mollie~E Schwartz, Jochen Braum{\"u}ller, Philip Krantz,
  Joel I-J Wang, Simon Gustavsson, and William~D Oliver.
\newblock Superconducting qubits: Current state of play.
\newblock {\em Annual Review of Condensed Matter Physics}, 11:369--395, 2020.

\bibitem{molmer1999multiparticle}
Klaus M{\o}lmer and Anders S{\o}rensen.
\newblock Multiparticle entanglement of hot trapped ions.
\newblock {\em Physical Review Letters}, 82(9):1835, 1999.

\bibitem{sorensen1999quantum}
Anders S{\o}rensen and Klaus M{\o}lmer.
\newblock Quantum computation with ions in thermal motion.
\newblock {\em Physical Review Letters}, 82(9):1971, 1999.

\bibitem{lloyd1996universal}
Seth Lloyd.
\newblock Universal quantum simulators.
\newblock {\em Science}, 273:1073--1078, 1996.

\bibitem{kitaev1995quantum}
A~Yu Kitaev.
\newblock Quantum measurements and the abelian stabilizer problem.
\newblock {\em arXiv preprint quant-ph/9511026}, 1995.

\bibitem{barahona1982computational}
Francisco Barahona.
\newblock On the computational complexity of {Ising} spin glass models.
\newblock {\em Journal of Physics A: Mathematical and General}, 15(10):3241,
  1982.

\bibitem{gottesman2009quantum}
Daniel Gottesman and Sandy Irani.
\newblock The quantum and classical complexity of translationally invariant
  tiling and hamiltonian problems.
\newblock In {\em 2009 50th Annual IEEE Symposium on Foundations of Computer
  Science}, pages 95--104. IEEE, 2009.

\bibitem{farhi2000quantum}
Edward Farhi, Jeffrey Goldstone, Sam Gutmann, and Michael Sipser.
\newblock Quantum computation by adiabatic evolution.
\newblock {\em arXiv preprint quant-ph/0001106}, 2000.

\bibitem{aspuru2005simulated}
Al{\'a}n Aspuru-Guzik, Anthony~D Dutoi, Peter~J Love, and Martin Head-Gordon.
\newblock Simulated quantum computation of molecular energies.
\newblock {\em Science}, 309(5741):1704--1707, 2005.

\bibitem{knill1997theory}
Emanuel Knill and Raymond Laflamme.
\newblock Theory of quantum error-correcting codes.
\newblock {\em Physical Review A}, 55(2):900, 1997.

\bibitem{kitaev2003fault}
A~Yu Kitaev.
\newblock Fault-tolerant quantum computation by anyons.
\newblock {\em Annals of Physics}, 303(1):2--30, 2003.

\bibitem{dennis2002topological}
Eric Dennis, Alexei Kitaev, Andrew Landahl, and John Preskill.
\newblock Topological quantum memory.
\newblock {\em Journal of Mathematical Physics}, 43(9):4452--4505, 2002.

\bibitem{bravyi1998quantum}
Sergey~B Bravyi and A~Yu Kitaev.
\newblock Quantum codes on a lattice with boundary.
\newblock {\em arXiv preprint quant-ph/9811052}, 1998.

\bibitem{freedman2001projective}
Michael~H Freedman and David~A Meyer.
\newblock Projective plane and planar quantum codes.
\newblock {\em Foundations of Computational Mathematics}, 1(3):325--332, 2001.

\bibitem{wang2003confinement}
Chenyang Wang, Jim Harrington, and John Preskill.
\newblock Confinement-{Higgs} transition in a disordered gauge theory and the
  accuracy threshold for quantum memory.
\newblock {\em Annals of Physics}, 303(1):31--58, 2003.

\bibitem{raussendorf2007fault}
Robert Raussendorf and Jim Harrington.
\newblock Fault-tolerant quantum computation with high threshold in two
  dimensions.
\newblock {\em Physical Review Letters}, 98(19):190504, 2007.

\bibitem{raussendorf2007topological}
Robert Raussendorf, Jim Harrington, and Kovid Goyal.
\newblock Topological fault-tolerance in cluster state quantum computation.
\newblock {\em New Journal of Physics}, 9(6):199, 2007.

\bibitem{Fowler_2013}
Austin~G. Fowler, Simon~J. Devitt, and Cody Jones.
\newblock Surface code implementation of block code state distillation.
\newblock {\em Scientific Reports}, 3(1):1--6, Jun 2013.

\bibitem{fowler2018low}
Austin~G Fowler and Craig Gidney.
\newblock Low overhead quantum computation using lattice surgery.
\newblock {\em arXiv preprint arXiv:1808.06709}, 2018.

\bibitem{litinski2019game}
Daniel Litinski.
\newblock A game of surface codes: Large-scale quantum computing with lattice
  surgery.
\newblock {\em Quantum}, 3:128, 2019.

\bibitem{litinski2019magic}
Daniel Litinski.
\newblock Magic state distillation: Not as costly as you think.
\newblock {\em Quantum}, 3:205, 2019.

\bibitem{horsman2012surface}
Clare Horsman, Austin~G Fowler, Simon Devitt, and Rodney Van~Meter.
\newblock Surface code quantum computing by lattice surgery.
\newblock {\em New Journal of Physics}, 14(12):123011, 2012.

\bibitem{gidney2021factor}
Craig Gidney and Martin Eker{\aa}.
\newblock How to factor 2048 bit {RSA} integers in 8 hours using 20 million
  noisy qubits.
\newblock {\em Quantum}, 5:433, 2021.

\bibitem{aharonov1996limitations}
Dorit Aharonov, Michael Ben-Or, Russell Impagliazzo, and Noam Nisan.
\newblock Limitations of noisy reversible computation.
\newblock {\em arXiv preprint quant-ph/9611028}, 1996.

\bibitem{zeng2019quantum}
Bei Zeng, Xie Chen, Duan-Lu Zhou, and Xiao-Gang Wen.
\newblock {\em {Quantum Information Meets Quantum Matter}}.
\newblock Springer, 2019.

\bibitem{ryu2006holographic}
Shinsei Ryu and Tadashi Takayanagi.
\newblock Holographic derivation of entanglement entropy from the anti--de
  {Sitter} space/conformal field theory correspondence.
\newblock {\em Physical Review Letters}, 96(18):181602, 2006.

\bibitem{weinberg1998}
Steven Weinberg.
\newblock Feynman on {God} and his granny; review of {The Meaning of It All},
  by {Richard Feynman}.
\newblock {\em The Times Higher Education Supplement}, (1334):23, 1998.

\bibitem{watson1958world}
Jane~Werner Watson.
\newblock {\em {The World of Science: Scientists at Work Today in Many
  Challenging Fields}}.
\newblock Golden Press, 1958.

\bibitem{sykes1981pleasure}
Christopher Sykes.
\newblock {The Pleasure of Finding Things Out}.
\newblock {\em Motion Picture}, 1981.

\bibitem{goodstein1989richard}
David~L Goodstein.
\newblock {Richard P. Feynman}, teacher.
\newblock {\em Physics Today}, 42(2):70, 1989.

\bibitem{feynman2010quantum}
Richard~P Feynman, Albert~R Hibbs, and Daniel~F Styer.
\newblock {\em {Quantum Mechanics and Path Integrals}}.
\newblock Courier Corporation, 2010.

\bibitem{baxter2016exactly}
Rodney~J Baxter.
\newblock {\em {Exactly Solved Models in Statistical Mechanics}}.
\newblock Elsevier, 2016.

\bibitem{feynman2018surely}
Richard~P Feynman.
\newblock {\em {Surely You're Joking, Mr. Feynman!: Adventures of a Curious
  Character}}.
\newblock WW Norton \& Company, 2018.

\bibitem{feynman2001you}
Richard~Phillips Feynman and Ralph Leighton.
\newblock {\em {What Do You Care What Other People Think?: Further Adventures
  of a Curious Character}}.
\newblock WW Norton \& Company, 2001.

\end{thebibliography}

\end{document}